\documentclass[draftclsnofoot, onecolumn]{IEEEtran}

\pdfoutput=1

\usepackage[dvips]{graphicx}
\usepackage{epstopdf}
\usepackage[]{caption}
\usepackage[caption=false,font=footnotesize]{subfig}
\usepackage{cite}
\usepackage{amsfonts}
\usepackage{amssymb}
\usepackage{amsmath}

\usepackage{xcolor}
\usepackage{xfrac}

\graphicspath{{.}}
\DeclareGraphicsExtensions{.pdf}

\newcommand{\BIBFILE}{../../bibtex/refs}

\begin{document}


\title{Algorithms for Embedding Quantum-Dot Cellular Automata Networks onto a Quantum Annealing Processor.}

\author{Jacob Retallick,
		Michael Babcock,
		Miguel Aroca-Ouellette,
		Shane McNamara,
		Steve Wilton,
		Aidan Roy,
		Mark Johnson,
		and Konrad Walus,~\IEEEmembership{Member,~IEEE}%
\thanks{This work was supported by the Natural Sciences and Engineering Research Council of Canada.}
\thanks{J. Retallick, M. Aroca-Ouellette, S. McNamara, S. Wilton, and K. Walus are with the Department of Electrical and Computer Engineering, University of British Columbia, Vancouver, BC V6T 1Z4, Canada (email:jret@ece.ubc.ca;  marocaou@caltech.edu; shanemcnamara11@gmail.com; stevew@ece.ubc.ca; konradw@ece.ubc.ca)}%
\thanks{M. Babcock, A. Roy, and M. Johnson are with D-Wave Systems Inc., Burnaby, BC V5G 4M9, Canada (email: michaelsbabcock@gmail.com; aroy@dwavesys.com; awjohnson@dwavesys.com)}
}


\maketitle

\begin{abstract}
Advancements in computing based on qubit networks, and in particular the flux-qubit processor architecture developed by D-Wave System's Inc., have enabled the physical simulation of quantum-dot cellular automata (QCA) networks beyond the limit of classical methods. However, the embedding of QCA networks onto the available processor architecture is a key challenge in preparing such simulations. In this work, two approaches to embedding QCA circuits are characterized: a dense placement algorithm that uses a routing method based on negotiated congestion; and a heuristic method implemented in D-Wave's Solver API package. A set of benchmark QCA networks is used to characterise the algorithms and a stochastic circuit generator is employed to investigate the performance for different processor sizes and active flux-qubit yields.
\end{abstract}



\section{Introduction}
\label{sec:intro}

Quantum-dot cellular automata (QCA) is a computational nanotechnology that employs a 2D patterned array of charge or magnetically coupled finite-state devices, QCA cells, for both classical and potentially quantum computing. There have been several experimental demonstrations of this technology in tunnel coupled metal-islands \cite{amlani1999}, patterned nanomagnets \cite{csaba2002}, and silicon-on-insulator \cite{macucci2003} including recent work in mixed-valence molecules \cite{lu2013, christie2015, tsuk2015} and atomic scale QCA built with charge coupled silicon dangling bond quantum dots \cite{wolkow2014}. Theoretical work on QCA was started by C.S. Lent \emph{et al,} in 1993 with a seminal paper introducing the 5-dot QCA cell and QCA ground state computing \cite{lent1993}. QCA networks have been predominantly treated using a Ising spin-glass model as it has been shown that the state of a properly operating binary QCA device remains relatively well described in the two-state approximation \cite{tougaw1996}. While a full quantum mechanical treatment scales as $2^N$, where $N$ is the number of cells, T\'oth \emph{et al.} showed theoretically that long-range correlations and entanglement between QCA cells was not necessary in reproducing QCA network dynamics \cite{toth01}.

QCA operation is fundamentally a quantum annealing process, with each component of a QCA circuit designed such that the ground state yields the desired output logic. Clocking is necessary in order to enforce directionality of information flow through the circuit and is achieved by controlling the cell parameters such that in the two-state approximation there is an effective modulation of the tunnelling between the polarization states \cite{hennessy2001}. So-called ``clocking-zones'' are sections of the circuit with cells assigned identical tunnelling in each clock phase. The cells in each subsequent clocking-zone are transitioned from their unpolarized ``relaxed'' state with strong tunnelling, to a polarized ``latched'' state with weak tunnelling. This is analogous to the usual quantum annealing procedure, in which a state is driven to its ground state by reducing the relative strength of the inter-state tunnelling. This suggests that some insight might be gained into the performance of QCA circuits by investigating an analogous quantum annealing architecture. In particular, such an architecture could probe the impact of coherent and environmental contributions to the dynamics during QCA clocking,

Work by D-Wave Systems Inc. in Burnaby Canada to realize a large scale quantum annealing platform \cite{dwave1,dwave2} has made it possible to explore the application of quantum annealing to the quantum simulation of large QCA circuits. The two-state approximation of QCA cell interactions and the interactions of qubits in D-Wave's quantum annealing processor (QAP) both produce governing Hamiltonians of analogous form. With a suitable embedding, the cell polarizations of a QCA circuit map directly to the expectations of qubit spins. However, the number of qubits on the QAP is limited and presently each qubit only couples to at most six other qubits. The sparse connectivity means that it is generally not possible to embed larger circuits using a one to one mapping between individual qubits and QCA cells. Additional qubits are needed to facilitate all interactions, with multiple qubits representing a single QCA cell. The choice of these multi-qubit groups defines the \emph{embedding} of a QCA circuit. It is not yet clear how these additional qubits contribute to the statistics of the quantum annealing process for a QCA circuit or what effect they have on the expected ground state.

In Section \ref{sec:theory}, we provide a brief overview of the theoretical principles of QCA cells and the flux-qubits as implemented in D-Wave's architecture. In particular, we comment on the analogous form of cell-cell and qubit-qubit interaction Hamiltonians in the two-state approximation. In Section \ref{sec:embedding}, we discuss the problem of embedding and present two approaches to finding QCA circuit embeddings. In Section \ref{sec:embedding-results}, we present embedding results using benchmarks circuits and use a stochastic circuit generator to estimate the range of circuits that can be successfully embedded onto different processor sizes, with a focus on the 512 qubit ``Vesuvius'' QAP. An earlier version of this paper was presented at COMMAD 2014 \cite{ret14}. In this expanded work, we discuss aspects of the algorithm implementations and extend our analysis to better characterise the embedding algorithms over a range of processor sizes and yields.


\section{Theoretical Background}
\label{sec:theory}

\begin{figure}
\centering
\newcommand{\ww}{.25\columnwidth}
\newcommand{\WW}{.29\columnwidth}
\subfloat[Ferromagnetic]{\raisebox{4ex}{\includegraphics[width=\ww]{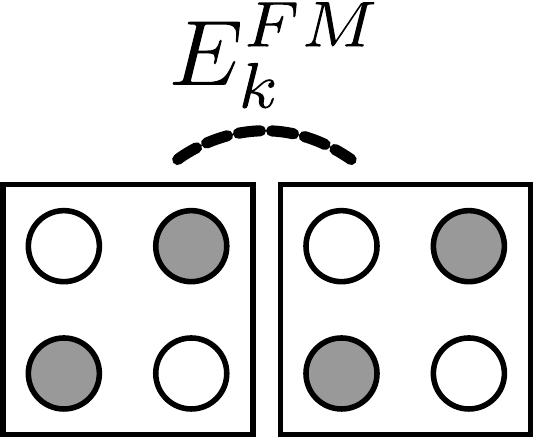}}} \qquad
\subfloat[Anti-ferromagnetic]{\includegraphics[width=\WW]{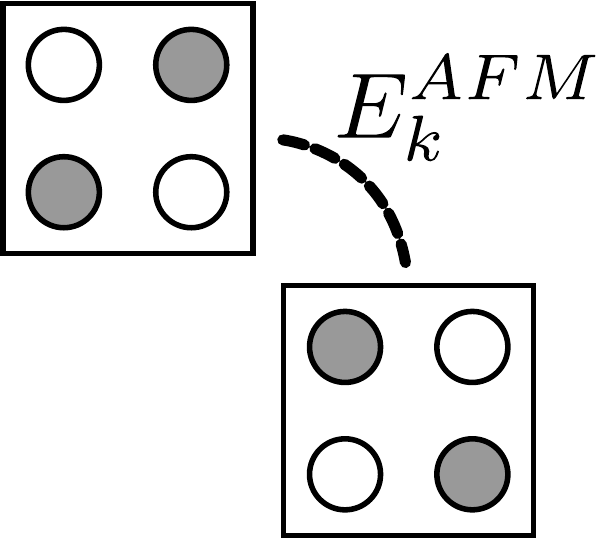}}
\caption{Typical QCA cell interactions with kink-energies. Ferromagnetically coupled cells tend to have the same polarization while anti-ferromagnetically coupled cells tend to have opposite polarization.}
\label{fig:cell-int}
\end{figure}

In the two-state approximation, cell-cell interactions are determined by an interaction energy, $E_k$, referred to as the \emph{kink energy}. The kink energy is defined as the change in electrostatic energy when two cells are switched from the same to opposite polarizations. This energy is a function only of the cell geometries and not the cell polarizations themselves. Typical arrangements for interacting 4-dot QCA cells are shown in Fig. \ref{fig:cell-int}. In keeping with the terminology used in Ising spin-glass models we use \emph{ferromagnetic} and \emph{anti-ferromagnetic} to refer to interactions that preferentially result in the same or opposite polarizations respectively. A layout of $N$ such QCA cells has an Ising-like Hamiltonian \cite{karim14},
\begin{eqnarray}
\label{eqn:two-state-Ham}
H_{QCA} = &-\sum_{i=1}^N \gamma_i \hat{\sigma}_x(i) - \frac{1}{2}\sum_{i<j}^N E_k^{i,j} \hat{\sigma}_z(i) \hat{\sigma}_z(j)\nonumber\\
&{+}\frac{1}{2}\sum_D\sum_{i=1}^N E_k^{i,D}P_D\hat{\sigma}_z(i),
\end{eqnarray}
where $\gamma_i$ is an effective tunnelling energy, $E_k^{i,j}$ is the kink energy between cells $i$ and $j$, $P_D$ is the polarization of driver $D$, $E_k^{i,D}$ is the kink energy between cell $i$ and driver $D$ and $\hat{\sigma}_\alpha(i)$ are the Pauli operators for the $i^{th}$ cell with $\alpha \in \{x,y,z\}$. For the configurations shown in Fig. \ref{fig:cell-int}, the kink energies with respect to an energy scale $E_0$ can be determined to be $E_k^{FM} = E_0$ and $E_k^{AFM} \approx -0.2 E_0$ for ferromagnetic and anti-ferromagnetic interactions respectively.

D-Wave's processor uses interacting superconducting flux qubits and has an operational Hamiltonian approximately of the form \cite{dickson2013}
\begin{eqnarray}
\label{eqn:D-Wave-Ham}
H_{DW}(t) = -\frac{1}{2}\sum_i \Delta_i(t) \hat{\sigma}_x(i) + \frac{1}{2} \mathcal{E}(t)\mathcal{H}_P,\\
\mathcal{H}_P = \sum_i h_i \hat{\sigma}_z(i) + \sum_{i<j} J_{ij} \hat{\sigma}_z(i)\hat{\sigma}_z(j),
\end{eqnarray}
where $\Delta_i(t)$ and $\mathcal{E}(t)$ are time dependent energy scales that define the \emph{annealing schedule} and $h_i$ and $J_{ij}$ are constants which describe a given Ising spin type optimization problem. Letting $t_f$ be the annealing time, the eigenspectrum after annealing can be made equal to that of a QCA circuit by setting
\[
\Delta_i(t_f) 	= 2\gamma_i, \quad
J_{ij} 	= -\frac{E_k^{i,j}}{\mathcal{E}(t_f)}, \quad
h_i		= \frac{\sum_{D}E_k^{i,D}P_D}{\mathcal{E}(t_f)}.
\]
The $h_i$ and $J_{ij}$ are dimensionless variables with values in the interval $[-1,+1]$. As the $\gamma_i$ are typically small for latched clocking zones, it is appropriate to allow $\Delta_i(t_f) \to 0$ which simplifies qubit readout. There is in general some error in assignment of $h_i$ and $J_{ij}$ values so it is preferable to scale the $E_k^{i,j}$ and $\sum_D E_k^{i,D} P_D$ to have maximum magnitude in order to minimize the relative error. In practice, this means scaling by $E_k^{FM}$ such that $J^{FM} = -1$ is the interaction strength between ferromagnetically coupled QCA cells.


\section{Embedding}
\label{sec:embedding}

Given a configuration of QCA cells, we aim to find a set of qubits and couplers such that the ground state achieved by quantum annealing maps to the ground state polarizations of the QCA circuit. In this section we discuss the convenience of representing QCA circuits as connectivity graphs, give an interpretation of the additional qubits necessary in embedding to facilitate the full set of interactions, describe the two embedding methods, and describe a method for converting the solutions to a form useful for parameter assignment and state readout after annealing.

\subsection{Graph Representation}
\label{subsec:graph-representation}

\begin{figure}[!t]
\centering
\includegraphics[width=.55\columnwidth]{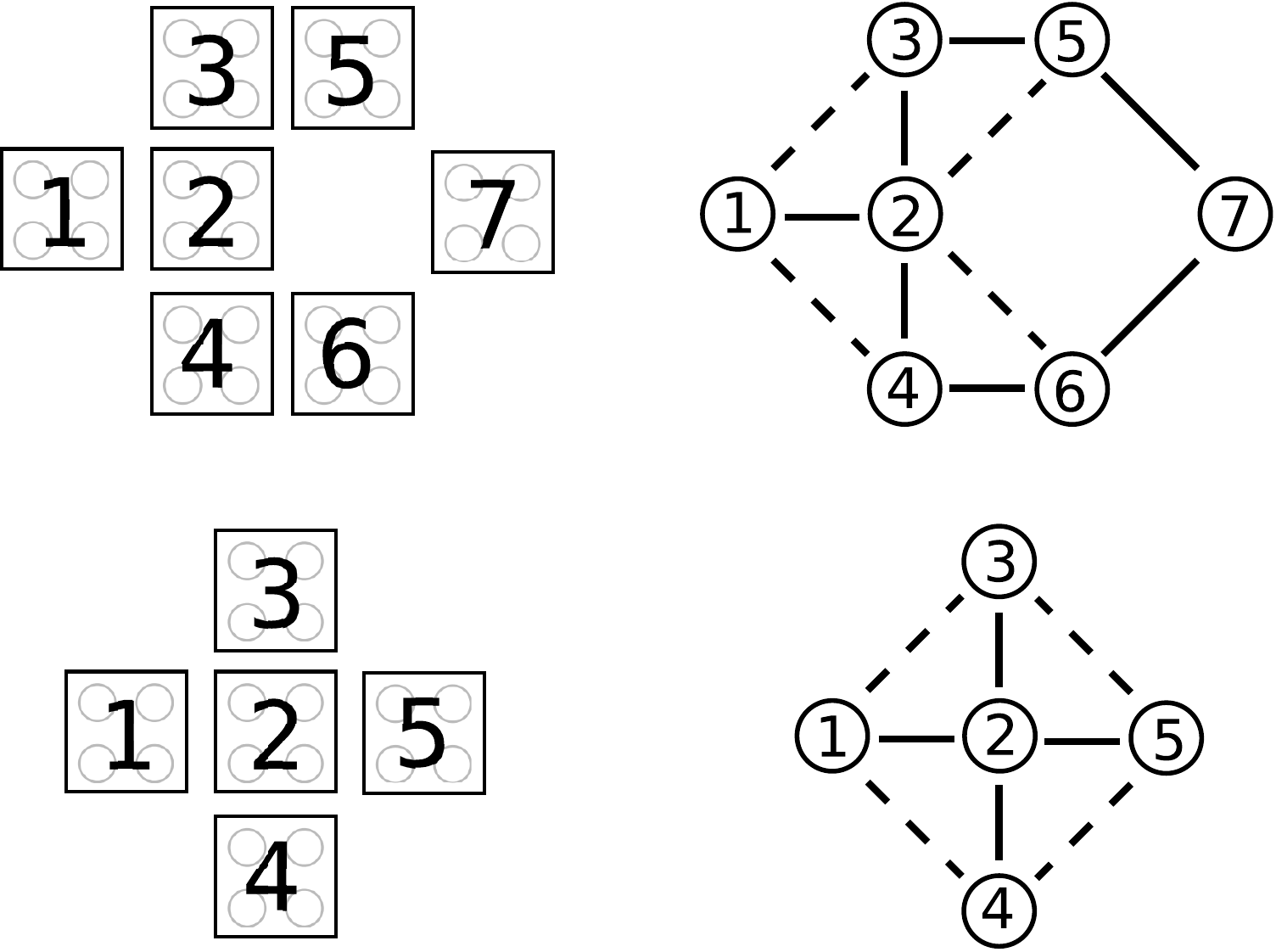}
\caption{Connectivity graph representations of an inverter (top) and majority gate (bottom). Coupling strengths are omitted for simplicity. Solid lines represent interactions considered for limited adjacency with dashed lines included for full adjacency.}
\label{fig:graph-rep}
\end{figure}

\begin{figure}[!t]
\centering
\subfloat[Qubit schematic]{\includegraphics[width=.38\columnwidth]{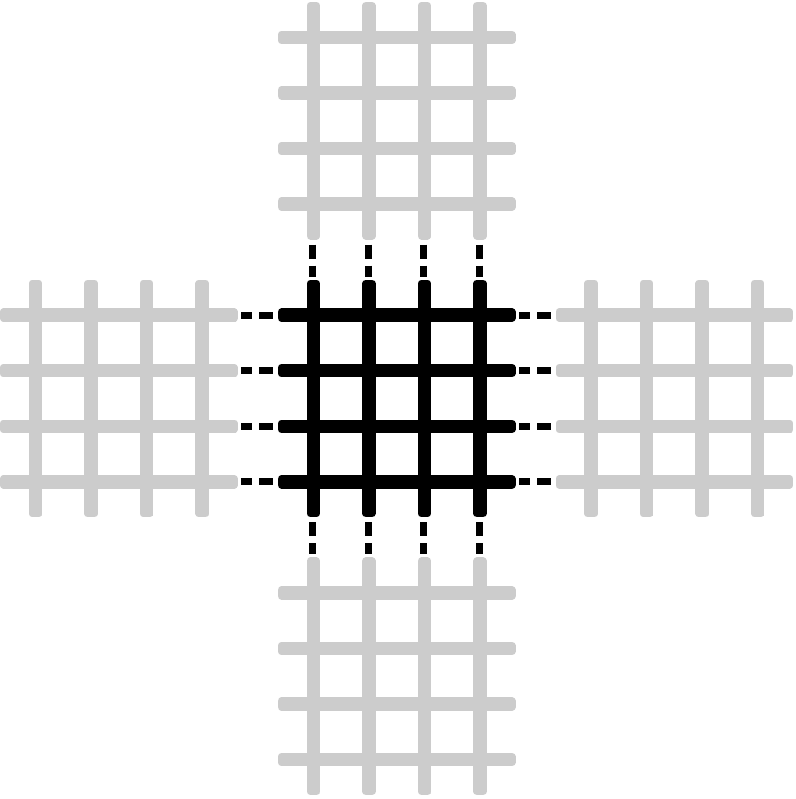}}\quad
\subfloat[Chimera graph]{\includegraphics[width=.38\columnwidth]{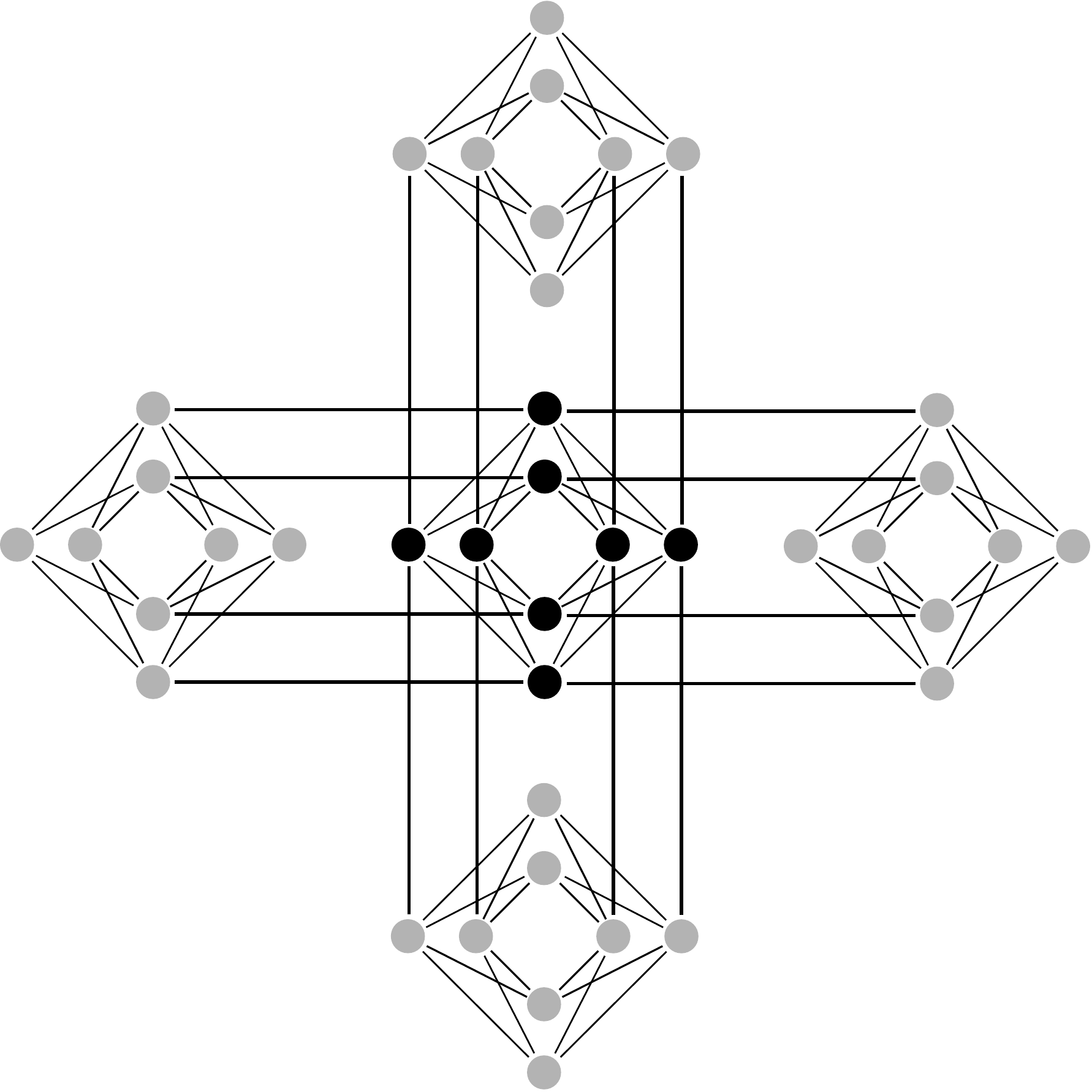}}
\caption{Qubit layout for one tile of the QAP. There are 4 vertical and 4 horizontal qubits with couplers at the intersections. Couplers between tiles are indicated by the dashed lines. In the graph representation qubits are represented by dots and couplers by connecting lines. Each qubit has 6 adjacent qubits (5 at the processor edge). D-Wave's Vesuvius processor has 512 qubits in an 8x8 array of tiles.}
\label{fig:chimera}
\end{figure}

For the purpose of consistent discussion, it is convenient to represent the cell interactions of a QCA circuit via a connectivity graph with driver cell contributions and cell-cell interactions mapping to node and edge weights respectively. As the quadrupole-quadrupole interaction between cells is approximately proportional to $r^{-5}$, where $r$ is the distance between cell centers, we consider only the interactions between cells within a radius of less than two cell sizes. The connectivity graph representations of the two fundamental QCA logic gates are shown in Fig. \ref{fig:graph-rep}. We define \emph{limited adjacency} as including diagonal (anti-ferromagnetic) cell interactions only in the case of an inverter and \emph{full adjacency} as when diagonal cell interactions are included for all cells. With this representation, the problem of embedding translates to finding the connectivity graph of the QCA circuit as a minor in the ``Chimera'' graph (single tile shown in Fig. \ref{fig:chimera}), which describes the available connections on D-Wave's QAP \cite{chimera}.

\subsection{Virtual QCA Cells}
\label{subsec:virtual-qca-cells}

\begin{figure}
\centering
\subfloat[Simple cell-cell interaction model]{\includegraphics[width=.4\columnwidth]{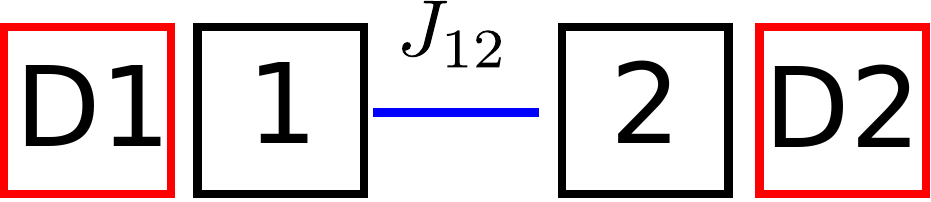}}\\
\subfloat[Virtual QCA cell group model]{\includegraphics[width=.8\columnwidth]{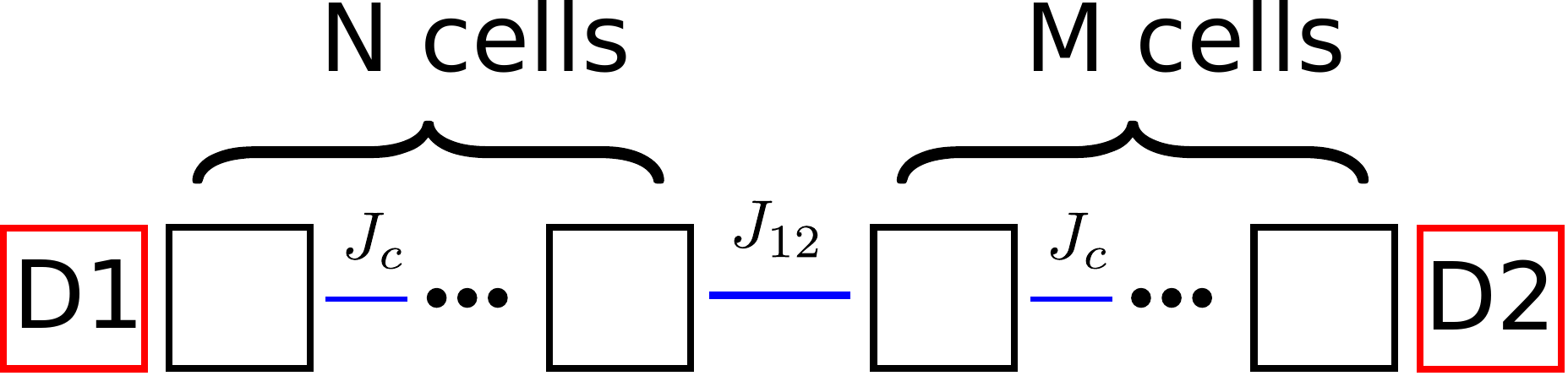}}
\caption{Cell-cell interaction model for estimating the effect of replacing cells by groups of virtual cells or qubits. Here the first and second QCA cells are replaced by wires of size $N$ and $M$ respectively.}
\label{fig:wire-int}
\end{figure}

\begin{figure}
\centering
\newcommand{\WW}{.42\columnwidth}
\subfloat[$J_{12} = 0.2$]{\includegraphics[width=\WW]{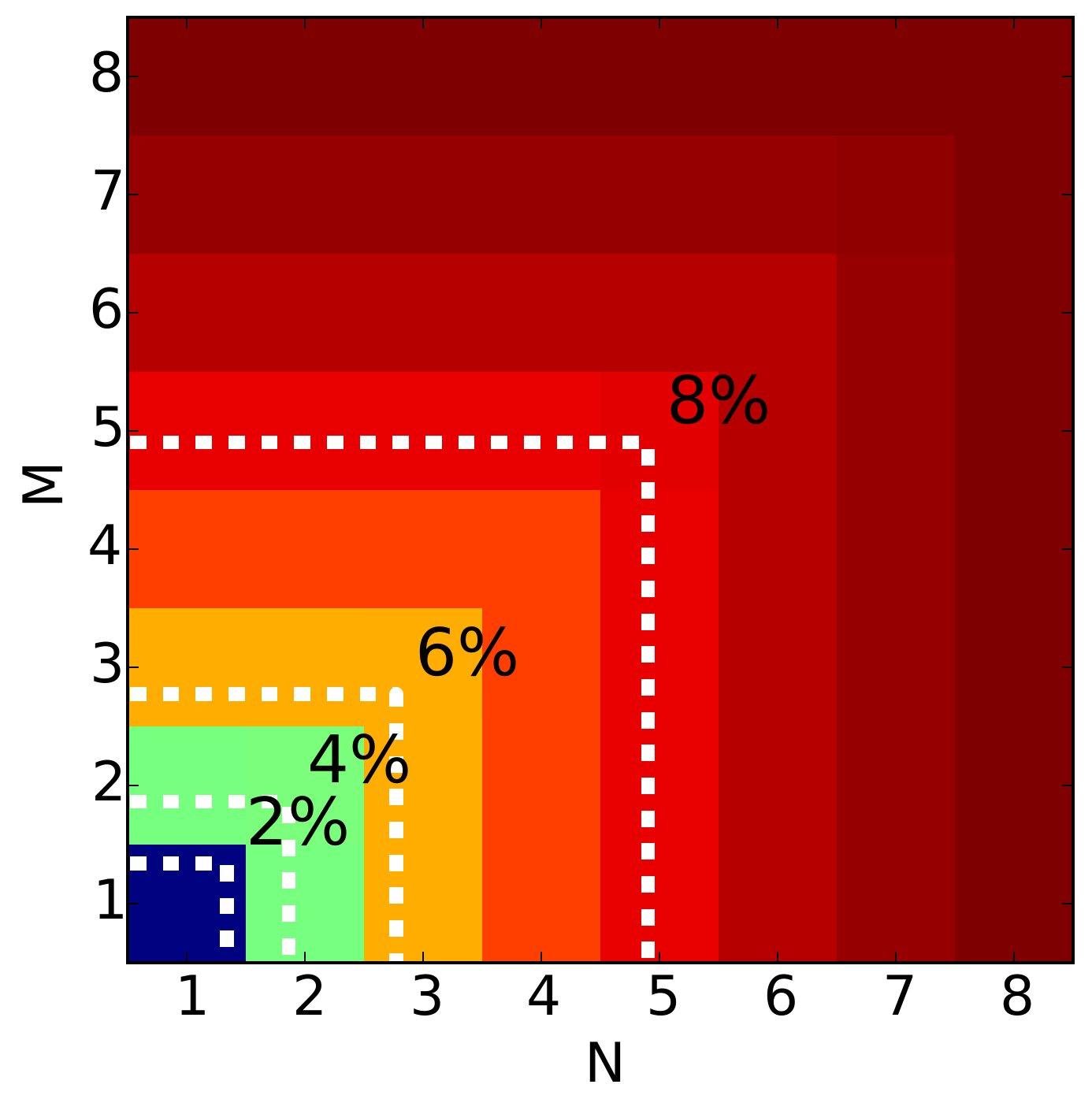}}\quad
\subfloat[$J_{12} = -1$]{\includegraphics[width=\WW]{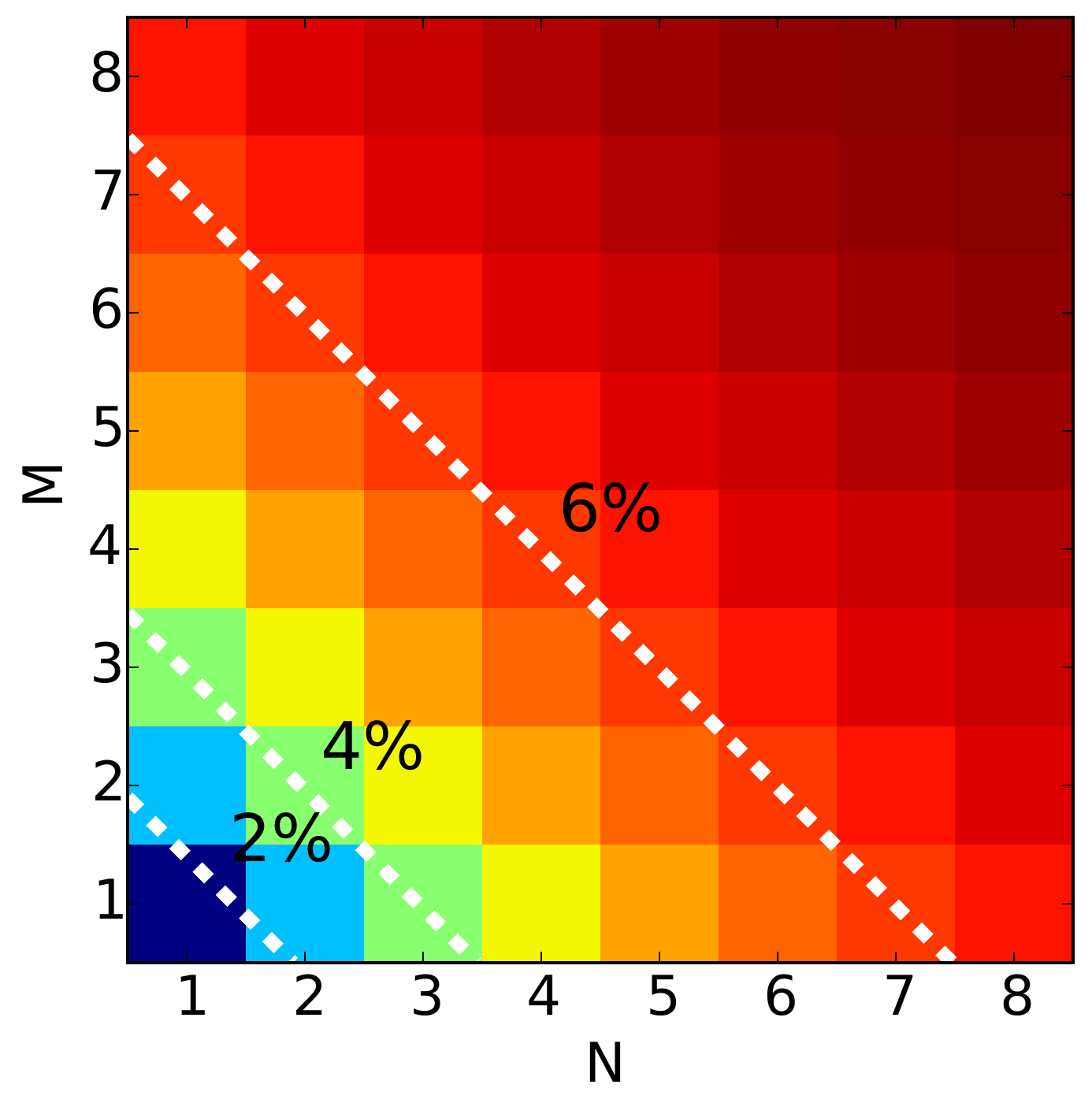}}
\caption{Dependence of the energy gap on the allocation of virtual qubits for $J_c = -1$. Dashed lines indicate 2\% decrements in the energy gap.}
\label{fig:wire_eg}
\end{figure}

Due to present constraints on connectivity, additional qubits are sometimes requires in order to accommodate all the relevant QCA network interactions. We can interpret these additional qubits within the framework of a QCA circuit as ``virtual'' QCA cells; these virtual cells, which do not exist in the original circuit, are inserted between ``real'' QCA cells to facilitate all original interactions.

As a simple model to understand the effects of these virtual cells or qubits we consider the single cell-cell interaction as shown in Fig. \ref{fig:wire-int}a. Here the driver cells on either side of the cells can represent an ICHA-like approximation of the coupling to the rest of the QCA circuit. The coupling strength between the two cells is indicated by $J_{12}$. We are interested in quantifying the effect of replacing each cell in the cell-cell interaction model by a group of coupled cells. Such a representation can be seen in Fig. \ref{fig:wire-int}b with internal coupling strength, $J_c$, between the cells inside each group as indicated.  $J_c$ will in general represent a strong ferromagnetic interaction.

As a metric for estimating the effect of such a replacement, we observe that in adiabatic annealing the probability of a non-adiabatic state evolution during an avoided crossing falls off as the minimum energy gap increases between the ground and first excited states \cite{landau, zener}. As adding more cells generally decreases the energy gap, we consider the percent reduction in the energy gap resulting from replacing the two cells by groups of size $N$ and $M$ respectively. In Fig. \ref{fig:wire_eg}, we consider these results both for the strong ferromagnetic ($J_{12} = -1$) and weak anti-ferromagnetic ($J_{12} \sim 0.2$) interactions common in QCA networks. Here we have used driver polarizations which would be expected for the given interactions: $P_{D1} = P_{D2} = 1$ for ferromagnetic coupling and $P_{D1} = -P_{D2} = 1$ for anti-ferromagnetic coupling. The stronger the internal coupling strength relative to the interaction strength between cell groups, the more accurately each cell group will match the behaviour of a single cell. As the parameter range for $h$ and $J$ is fixed, the strongest interaction we can achieve is $J_c = -1$. An effectively stronger $J_c$ can be achieved by reducing all $h$ and inter-group $J$ parameters by a constant factor however this would increase the relative error arising from parameter assignment accuracy on the processor. Observe that for $J_c = J_{12} = -1$ the boundary between the two groups is arbitrary and thus the energy gap can depend only on the total number of cells. If the coupling strength is weaker than the internal coupling as in $J_{12} = 0.2$ we observe that the energy gap approximately depends only on the maximum group size. As a general rule for either case, it is preferable to arrange the allocation of virtual cells or qubits in order to minimize the largest group size. The simulated percent decrease in the energy gap remained less than 10\% for group sizes up to 8 cells (16 cells total). These simple considerations suggest a minimal effect for small group sizes. Other effects such as interactions with the environment will likely make large virtual cell groups not perform as expected but we assume such effects are similarly limited for sufficiently small maximum group sizes.

\subsection{Dense Placement Embedding Algorithm}
\label{subsec:dense-placement-embedding}

The Dense Placement algorithms uses a simultaneous ``place-and-route'' approach commonly used in FPGA design \cite{mcmurchie95}. It attempts to place every node in the QCA connectivity graph onto a corresponding \emph{assigned qubit} in the Chimera graph. Interactions between cells are facilitated either by direct coupling or by chains of qubits routed between these assigned qubits.

\subsubsection{Initial Seed}

Before the main loop can be called, an initial seed must first be selected. The first QCA cell node and its corresponding qubit are chosen as follows. Each QCA cell $i$ is assigned a probability proportional to $A_i^p$ where $A_i$ is the number of neighbours of cell $i$ and $p$ is a constant. In this paper, $p$ was selected to be 3 in order to prioritize high adjacency cells but still allow some chance for lower adjacency cells to be selected. In selecting the seed qubit, each tile is assigned a probability according to a Gaussian distribution about the center of the tile array with width $\sigma$. The qubits of that tile are then randomly ordered and iterated through until a qubit is found with enough neighbours to facilitate the interactions of the selected seed cell. If no such qubit is found, a new tile is selected. In this work, $\sigma$ was chosen to be 1 to strongly bias initial placement near the center of the Chimera graph.

\subsubsection{Cell Placement}

Building from the initial seed, for each iteration of the main loop we look at the set of unplaced cells adjacent to the set of placed cells and sort by decreasing number of unplaced connections. For each cell to be placed, we aim to select a qubit with sufficiently many available neighbouring qubits to facilitate all interactions. Such a qubit is referred to as \emph{suitable}. The chosen qubit is that which is both suitable and yields the shortest total path to all qubits assigned to adjacent cells. This is achieved by simultaneously running a lowest cost path search from each such qubit until a suitable qubit is reached by all search trees. The cost of each edge in the search contains two components: a contribution from the type of coupling to the new node (internal or external to a tile), and an edge proximity contribution which weighs nodes in tiles closer to the edge of the processor more strongly. For dense placement, internal tile placements are preferred so internal couplers have lower cost than external couplers. The external coupler cost was chosen to be $1.9$ times the internal coupler cost such that one external coupler was preferable to two internal couplers. The edge proximity cost increases linearly as the minimum distance to the processor edge decreases. An increase in cost of 0.5 per tile was chosen for the 512 qubit processor. These parameters were chosen to tightly pack long chains of low adjacency cells and avoid running out of room for embedding.

\subsubsection{Routing}

The routing algorithm is based on negotiated congestion, a well established path finding technique used in mapping netlists to FPGAs \cite{mcmurchie95}. This method aims to minimize the path length for each route by use of a cost scheme that penalizes the use of a qubit for more than one path.

\subsubsection{Seam Opening}

In the case that there is no suitable qubit available, the translational symmetry of the chimera graph allows an avenue of qubits and couplers to be ``opened'' by shifting all qubits after or before a row or column. We interpret this as opening a seam between two rows or columns. As specific qubits or couplers may be inactive due to manufacturing yield, some shifted qubits may require remapping. Such qubits are said to ``conflict''. Following such a shift, remapping is achieved by finding new placements for cells with conflicting assigned qubits and rerouting conflicting paths. The new cell placements can result in recursive calls to seam opening and remapping. A basic metric for selecting which seam to open and in which direction is used. For the given cell to be placed, we consider all seams adjacent to assigned qubits of the cell's placed neighbours. For each of these seams and each opening direction, we check that there is a free row or column to shift into. If so, we assign to that seam a cost, $C$, given by

\begin{equation}
C = c_q n_q + c_p n_p + c_d d ,
\end{equation}

where $n_q$ and $n_p$ are the number of conflicting assigned qubits and paths, $d$ is the seam distance defined as the minimum distance between the seam and the mean location of all assigned qubits, and $c_q$, $c_p$, and $c_d$ are the relevant cost factors. In this work, the cost factors were chosen as $c_q = 3$, $c_p = 4$, and $c_d = 3$. Typically, fewer qubits were needed for replacing a small number of qubits than re-routing a number of paths.

\begin{figure}
\centering
\includegraphics[width=.9\columnwidth]{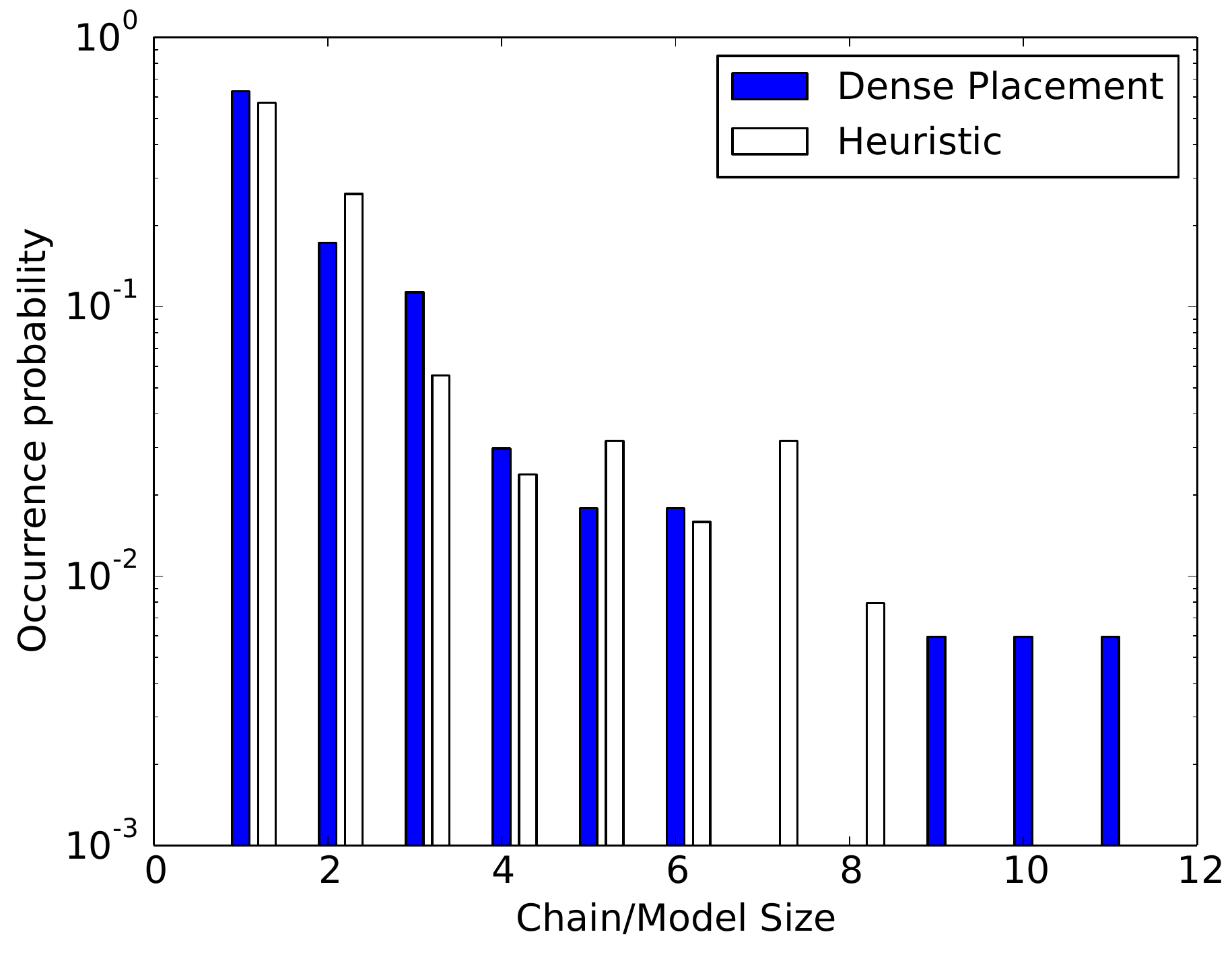}
\caption{Occurrence probability of qubit chain lengths and vertex-model sizes for the Dense Placement and Heuristic algorithms respectively for a sample trial of the serial adder with full adjacency. Chain lengths are measured by the number of couplers used with models measured by the number of qubits. Due to the different interpretations of embedding used by the two algorithms the distributions represent how each model allocates virtual qubits. Note the occurrence of long qubit chains for the Dense Placement algorithm.}
\label{fig:ser-add-dist}
\end{figure}

\begin{figure}
\centering
\newcommand{\HH}{2.5in}
\subfloat[Serial adder QCA cell layout. Driver cells are shaded.]{\includegraphics[height=\HH]{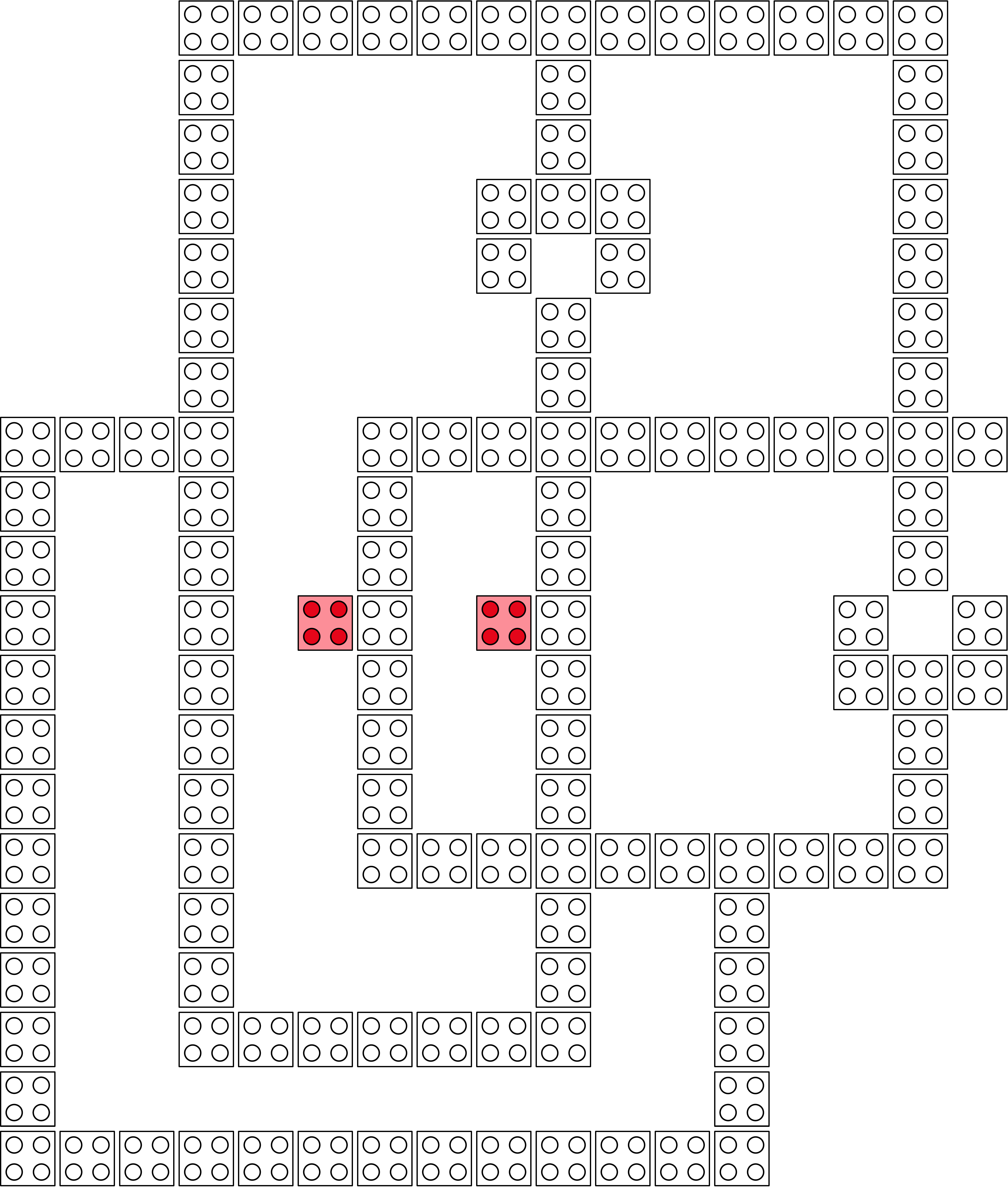}} \hspace{2cm}
\subfloat[Sample Embedding]{\includegraphics[height=\HH]{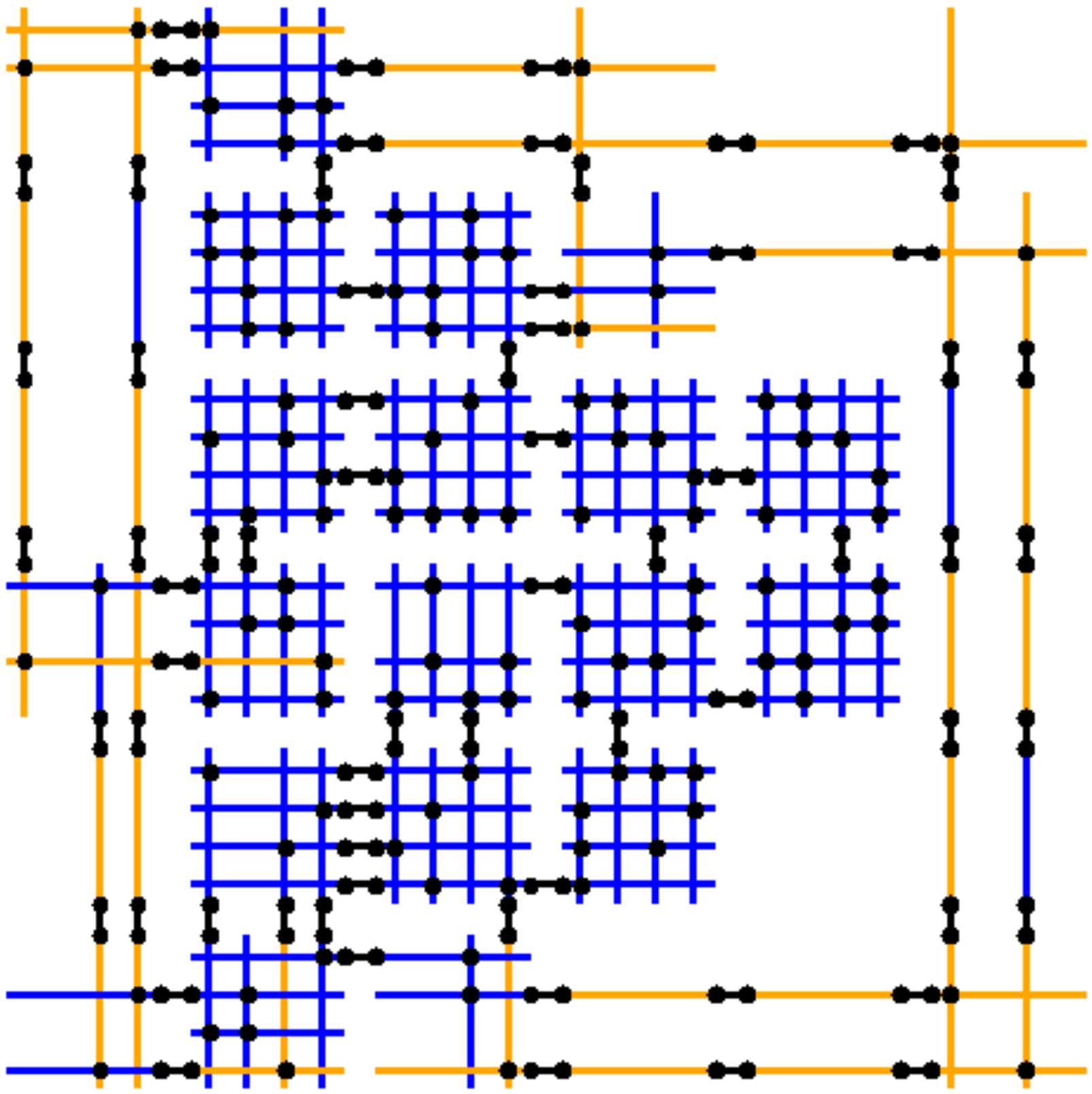}}
\caption{Example embedding of a serial adder QCA circuit. Here qubits are represented by line segments with active couplers shown by black circles. Darker shaded qubits indicate those assigned to QCA cells with lighter shaded qubits indicating virtual qubits. Only tiles with used qubits are shown.}
\label{fig:ser-add-emb}
\end{figure}

The Dense Placement algorithm is good at keeping most routes between mapped qubits to only one or two couplers. However, as the placement builds outwards from the center and does not currently account for how the final cells to be placed are connected, there are often long chains of ten or more qubits about the perimeter of the embedding needed to facilitate the final connections. Fig. \ref{fig:ser-add-dist} shows an example distribution of qubit chain lengths for an embedding of a serial adder. The QCA cell layout and a sample embedding for the serial adder are shown in Fig. \ref{fig:ser-add-emb}. Note the long chains about the perimeter as discussed.

\subsection{D-Wave's Heuristic Embedding Algorithm}
\label{subsec: d-waves-heuristic-embedding-algorithm}

D-Wave's Solver API package includes an embedding solver which uses a heuristic method to attempt to find a source graph as a minor in the hardware connectivity graph. For each node in the source graph, the heuristic algorithm attempts to find a connected sub-graph or ``vertex-model'' comprised of strongly coupled qubits. Edges between nodes in the source graph translate to shared edges between vertex-models in the Chimera graph. A detailed description of this algorithm can be found elsewhere \cite{cai2014} but the basic approach is provided here. 

For each node in the source graph, assign an initial vertex-model as follows. If none of the current node's neighbours have been assigned a vertex-model, randomly select a qubit in the target graph as the vertex-model. Otherwise, find the qubit with the smallest total path cost to all of the adjacent vertex-models and assign this qubit as the root of the vertex-model. The same qubit can be used by multiple paths at a high cost. Paths to the adjacent vertex-models are distributed between vertex-models to minimize the largest vertex-model size. After initial placement, iteratively loop through each source node, forget the previously assigned vertex-model and try to find a better one. The metrics for improvement are defined as the number of times a single qubit is used in the set of all vertex-models (must be no more than once for a successful embedding), the sum of vertex-model sizes, and the largest vertex-model size.

This method is good at reducing the total number of qubits used and the largest group of qubits assigned to a single QCA cell. Refer to Fig. \ref{fig:ser-add-dist} for an example of the distribution of vertex-model sizes. As the number of short vertex-models are not considered, fewer small groups are typically found than short chains in the Dense Placement algorithm.

\subsection{Parameter Assignment and Model Conversion}
\label{subsec:model-conversion}

Once an embedding is found, it is necessary to assign bias and coupler parameters $h_i$ and $J_{ij}$ to the included qubits and couplers respectively. In the vertex-model representation, this is easily achieved. The vertex-model for QCA cell $i$ has a set of qubits, $Q_i$, connected by internal couplers and a set of external couplers which couple $Q_i$ to each adjacent QCA cell vertex-model. The biases, $h_n$,  of the vertex-model must sum to the bias for the QCA cell. We distribute the bias uniformly as
\begin{equation}
h_n = \frac{1}{E_k^{FM} |Q_i|} \sum_D E_k^{i,D} P_D ,  \quad \forall n \in Q_i.\\
\end{equation}
The couplers between qubits within each model are all assigned the maximum ferromagnetic parameter $J_c = -1$. Typically there is only one coupler between each pair of vertex-models to which we assign the corresponding $J_{ij} = -\sfrac{E_k^{i,j}}{E_k^{FM}}$. If multiple couplers exist between two models, we can choose to uniformly distribute $J_{ij}$.

\begin{figure}
\centering
\includegraphics[width=.8\columnwidth]{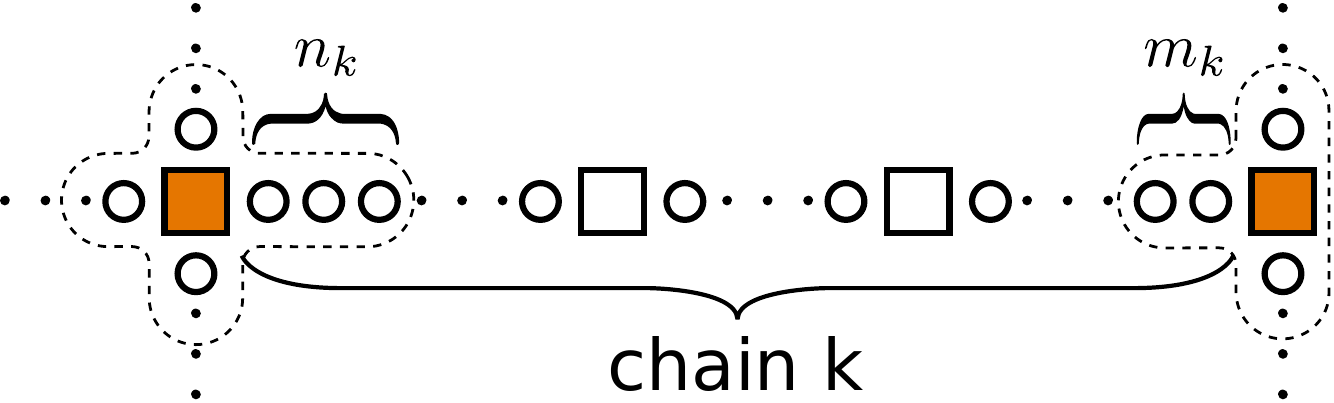}
\caption{Schematic of a chain between two end-nodes. Squares and circles represent assigned and virtual qubits respectively. The filled squares are the end-nodes terminating chain $k$ and the unfilled squares are internal-nodes within chain $k$. Dotted boundaries indicate the end-node vertex-models.}
\label{fig:model-rep}
\end{figure}

In the place-and-route representation, it is not so trivial to choose a scheme for parameter assignment. Between each pair of qubits assigned to adjacent QCA cells there can exist a chain of virtual qubits. The interaction term $J_{ij}$ between the two cells can be assigned to any coupler along this chain without changing the effective interaction between the assigned qubits, given that all other connectors are assigned $J_c=-1$. The act of parameter assignment then takes the form, using the terminology so-far defined, of finding an effective vertex-model for each cell by deciding where along each virtual qubit chain to assign the interaction term. Once the vertex-models are found, we can proceed as above to assign bias and interaction parameters.

In Sec. \ref{subsec:virtual-qca-cells}, it was suggested that the behaviour of two interacting vertex-models is optimized by minimizing the maximum vertex-model size. We assume this to be a general feature of any interacting network of vertex-models.  One can formulate the allocation of virtual qubits with the aim to minimize the maximum of all vertex-model sizes as a linear optimization problem. First, in order to maintain the correct number of connectors between vertex-models observe that any cell with greater or less than 2 neighbours must contain in its vertex-model its assigned qubit. It is then useful to reinterpret the QCA circuit connectivity graph as a set of ``end-node'' ($A_i \neq 2$) cells connected by chains of ``internal-node'' ($A_i=2$) cells. The Dense Placement algorithm then effectively constructs chains containing virtual and internal-nodes between the end-nodes. One such chain is illustrated in Fig. \ref{fig:model-rep}. Note that the internal-nodes could be placed anywhere along these chains without changing the embedding as long as their order is maintained. The problem of qubit allocation then reduces to determining the number of qubits at the end of each chain assigned to the vertex models of the end-nodes. The remaining qubits of each chain are divided evenly between the internal-node cells. Only chains containing virtual qubits need to be considered.

The $k^{th}$ chain contains $N_k$ internal-nodes and total of $M_k$ qubits, excluding the end-nodes. Of these qubits, $n_k$ and $m_k$ qubits are assigned to the vertex-models of the end-nodes at the beginning and end of the chain respectively. Then the $k^{th}$ chain has an average internal-node vertex-model size, $\bar{s}_k$, of
\begin{equation}
\bar{s}_k = \frac{M_k-n_k-m_k}{N_k}.
\end{equation}
If $S_\ell$ is the sum of all $n_i$, $m_j$ associated with the $\ell^{th}$ end-node, then each end-node has a vertex-model size, $s_\ell$ of
\begin{equation}
s_\ell = 1+S_\ell.
\end{equation}
The linear programming problem is then given as minimizing the maximum of the set of all $\bar{s}_k$ and $s_\ell$ with constraints $n_i, m_i \geq 0 \: \forall i$ and $n_i + m_i \leq M_i - N_i \: \forall i$. This is an integer programming problem and hence is potentially computationally intensive. Typically the number and allowed range of parameters is small so this was not observed to be a concern. If necessary, linear programming (LP) relaxation can be used, removing the integer constraint on $n_i$ and $m_i$, with some consideration to rounding in post-processing. This method optimizes the maximum vertex-model size but does not consider the distribution of the remaining vertex-models. Better results could be obtained from a non-linear optimization which considers all vertex-model sizes.


\section{Embedding Results}
\label{sec:embedding-results}

We ran both algorithms on a variety of circuits to determine the number of qubits required for a given number of QCA cells. We used a set of benchmark circuits to investigate the performance on designed QCA circuits as well as implemented a stochastic circuit generator to gain insight into the range of embeddable circuits. In sections \ref{subsec:benchmark-circuits} through \ref{subsec: scaling-performance}, we assume no disabled qubits or couplers on the target processor.

\subsection{Benchmark Circuits}
\label{subsec:benchmark-circuits}

\begin{table}
\renewcommand{\arraystretch}{1.3}
\caption{Benchmark circuits and the number of non-driver cells}
\label{table:benchmark-circuits}
\centering
\begin{tabular}{|c|c||c|c|}
\hline
\bfseries Circuit & \bfseries Cells & \bfseries Circuit & \bfseries Cells\\
\hline
SR Flip Flop	&20&Memory Loop&	120\\
Selectable Oscillator&31&Serial Adder&126\\
XOR Gate&73&4 Bit 2-1 Mux&192\\
Full Adder&98&4 Bit Accumulator&273\\
\hline
\end{tabular}
\end{table}

\begin{figure}
\centering
\newcommand{\WW}{.85\columnwidth}
\subfloat[Full Adjacency]{\includegraphics[width=\WW]{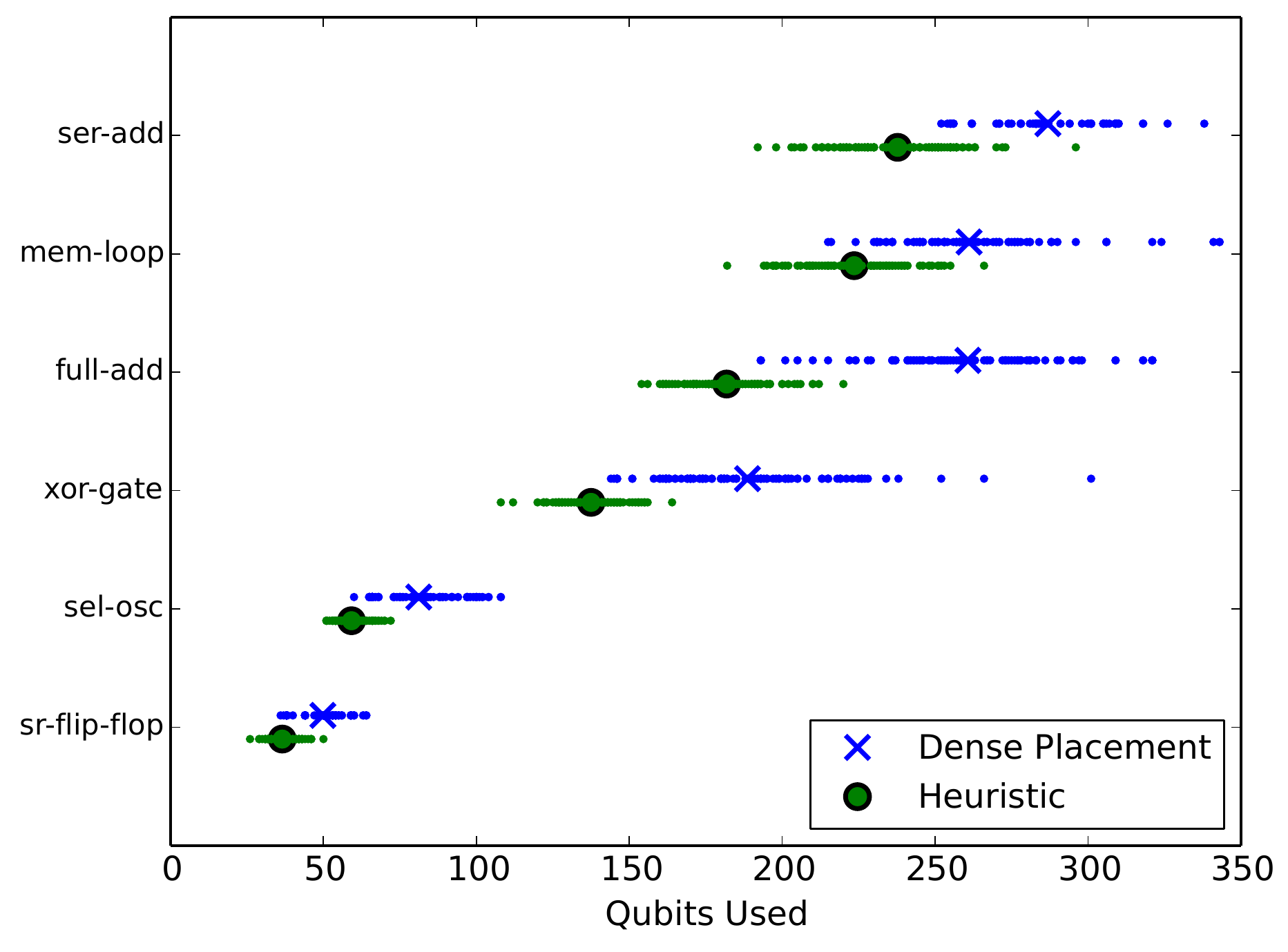}} \\
\subfloat[Limited Adjacency]{\includegraphics[width=\WW]{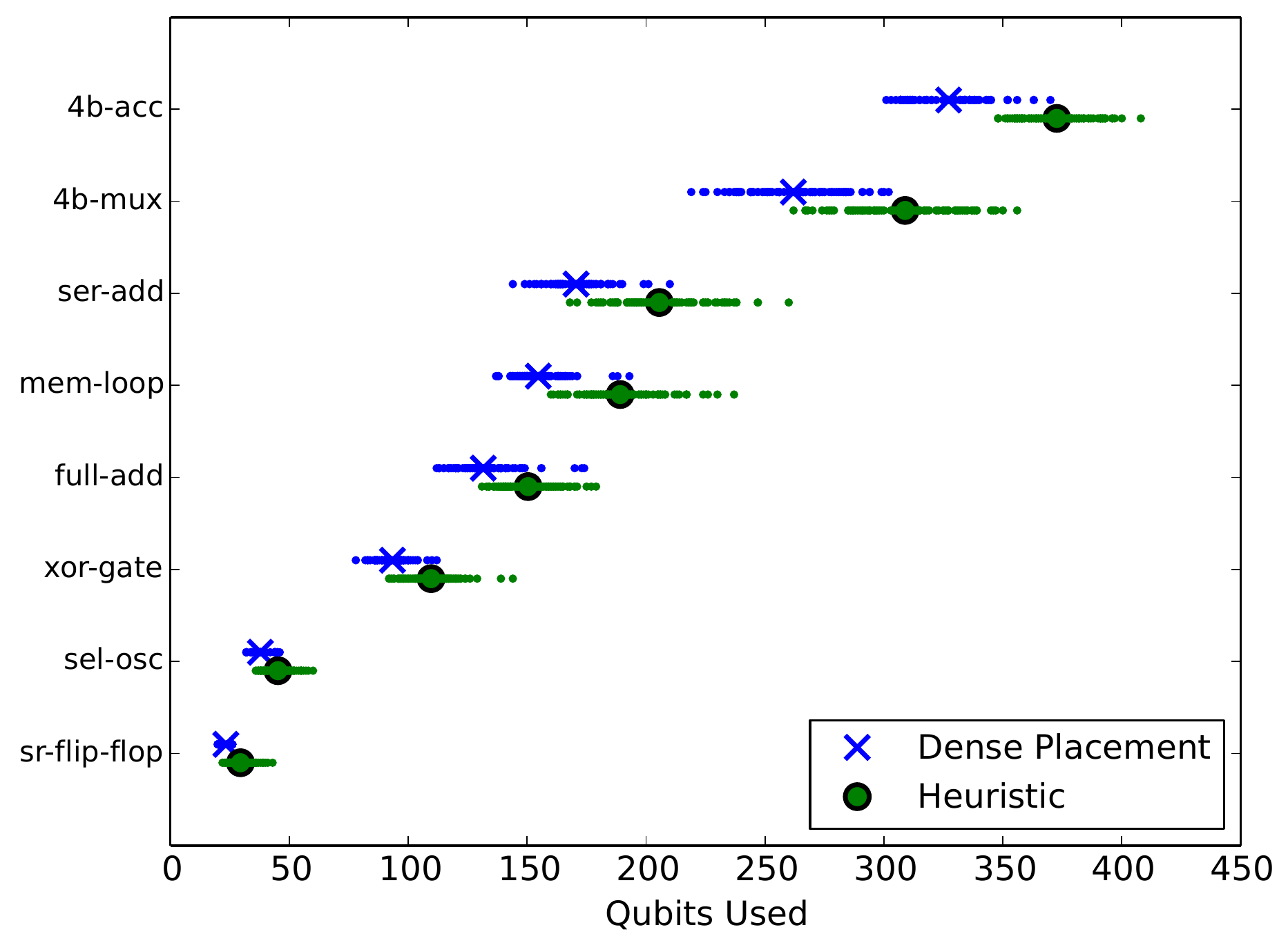}}
\caption{Number of qubits used for each benchmark circuit. The larger markers correspond to the average number of qubits used with each dot a successful trial embedding}
\label{fig:bench-embeds}
\end{figure}

\begin{figure}
\centering
\subfloat[Average qubit usage]{\includegraphics[width=.8\columnwidth]{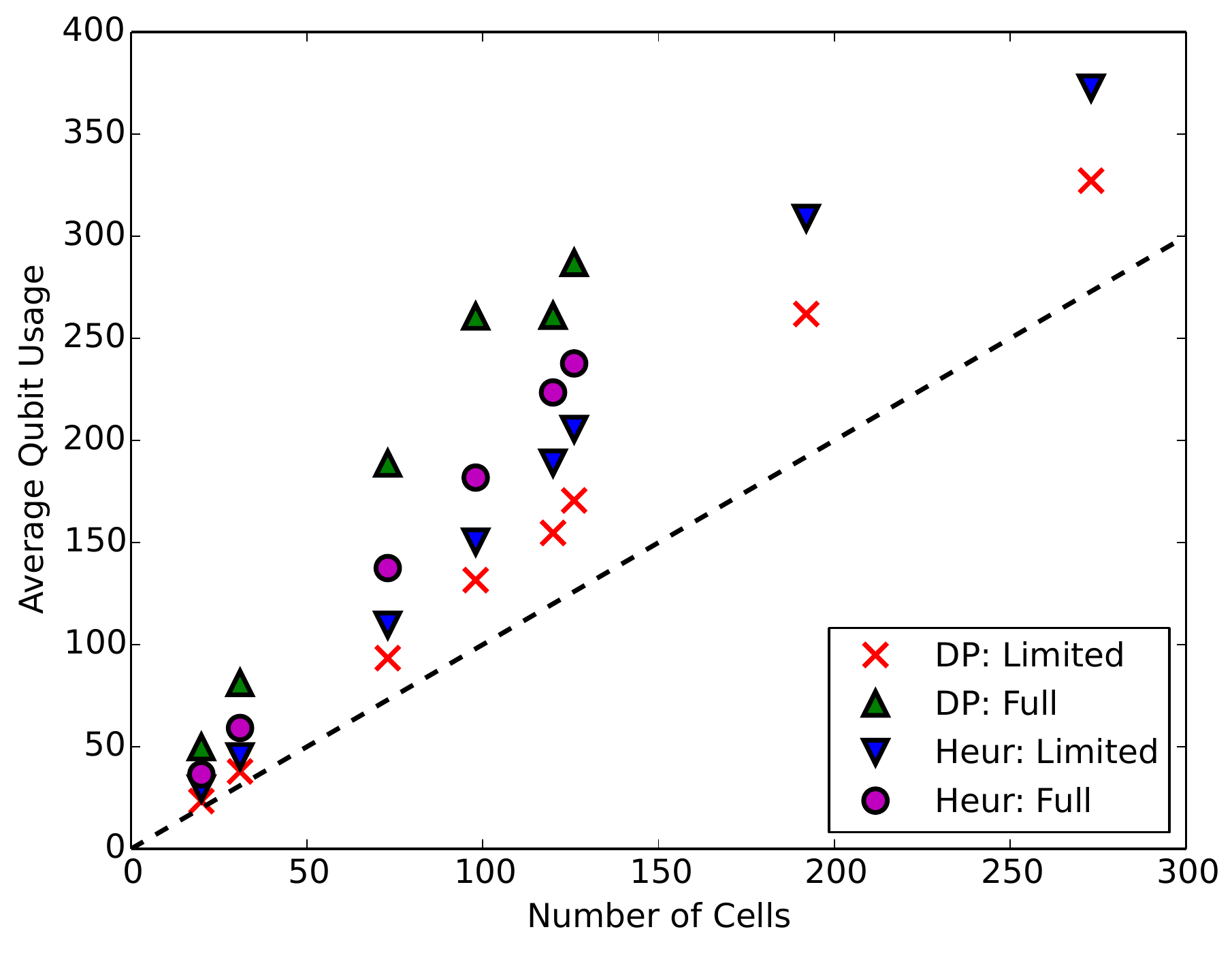}}\\
\subfloat[Average maximum model size]{\includegraphics[width=.8\columnwidth]{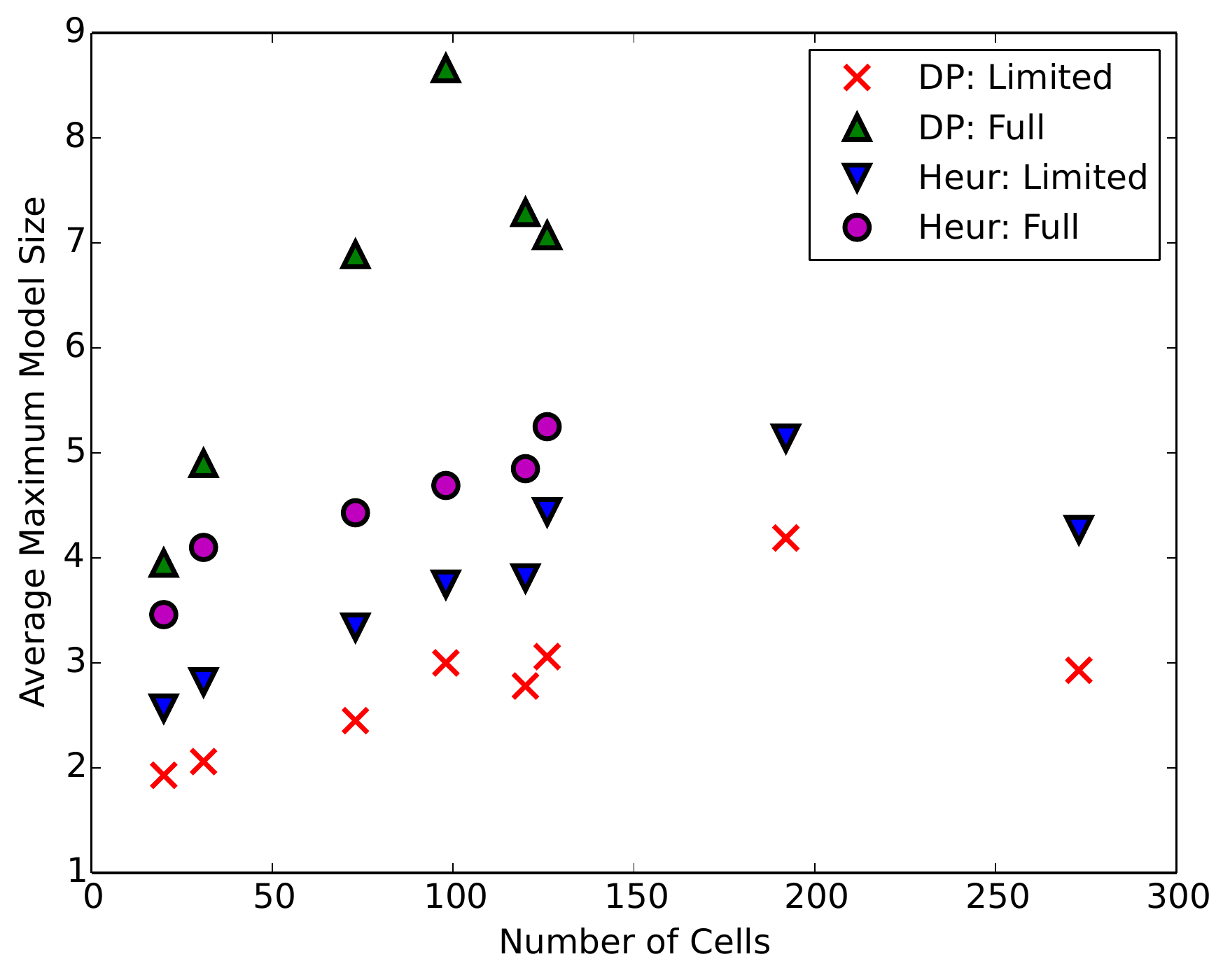}}
\caption{Benchmark Circuits: Dense Placement (DP) and Heuristic (Heur). Adjacency as indicated. The dashed line in (a) represents an ideal one-to-one embedding.}
\label{fig:summ}
\end{figure}

There were eight benchmark circuits used to test the embedding capabilities of the Dense Placement and Heuristic algorithms. The circuit names and numbers of non-driver cells are shown in Table \ref{table:benchmark-circuits}. None of the benchmark circuits contain crossover elements and the clocking arrangements were ignored. For each benchmark circuit, we attempted to find 100 embeddings using each algorithm for both limited and full adjacency. Fig. \ref{fig:bench-embeds} shows the number of qubits needed for each circuit for both adjacency types.

We observed that for full adjacency the Heuristic algorithm required fewer qubits whereas the Dense Placement algorithm required fewer qubits for limited adjacency. We also see that for each of the benchmark circuits, the Heuristic algorithm had a smaller reduction in qubit usage from full adjacency to limited adjacency than observed in the Dense Placement algorithm. This suggests that the Heuristic algorithm is much better than the Dense Placement algorithm at handling highly connected nodes but is not particularly efficient at embedding large numbers of sparsely connected nodes as is the case for limited adjacency. In contrast, the Dense Placement algorithm is good at tightly packing cells which have only one unplaced connection. In such cases it prioritizes same tile placement and hence condenses long wires into small regions of the Chimera graph. For comparison, a summary of the average qubit usages for the benchmark circuits for both methods and adjacencies is included in Fig. \ref{fig:summ}a. In all cases, the qubit usage seems to be approximately linear with circuit size. Fig. \ref{fig:summ}b. shows the average maximum  vertex-model size with conversion for the Dense Placement results achieved as discussed in Sec.  \ref{subsec:model-conversion}. While the maximum model size tends to increase slightly with the number of cells it seems to be more a function of the complexity of the circuit. This can be seen with the 4-bit accumulator circuit which has the most cells, most of which exist in long wires, but has a lower maximum model size than many of the smaller circuits.

\subsection{Generated Circuits}
\label{subsec:generated-circuits}

\begin{figure}
\centering
\newcommand{\WW}{.8\columnwidth}
\subfloat[Generated Circuits: Dense Placement Algorithm]{\includegraphics[width=\WW]{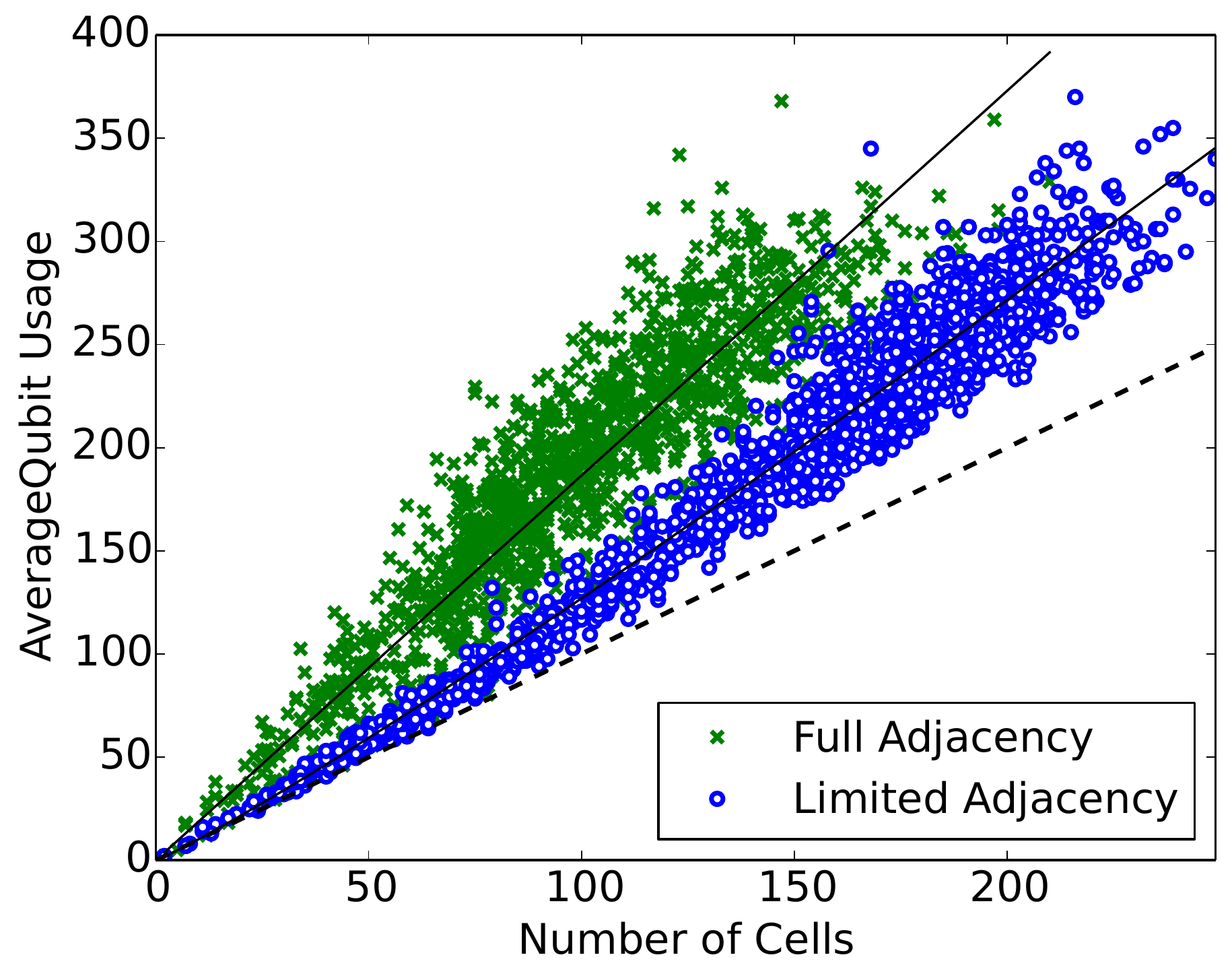}}\\
\subfloat[Generated Circuits: Heuristic Algorithm]{\includegraphics[width=\WW]{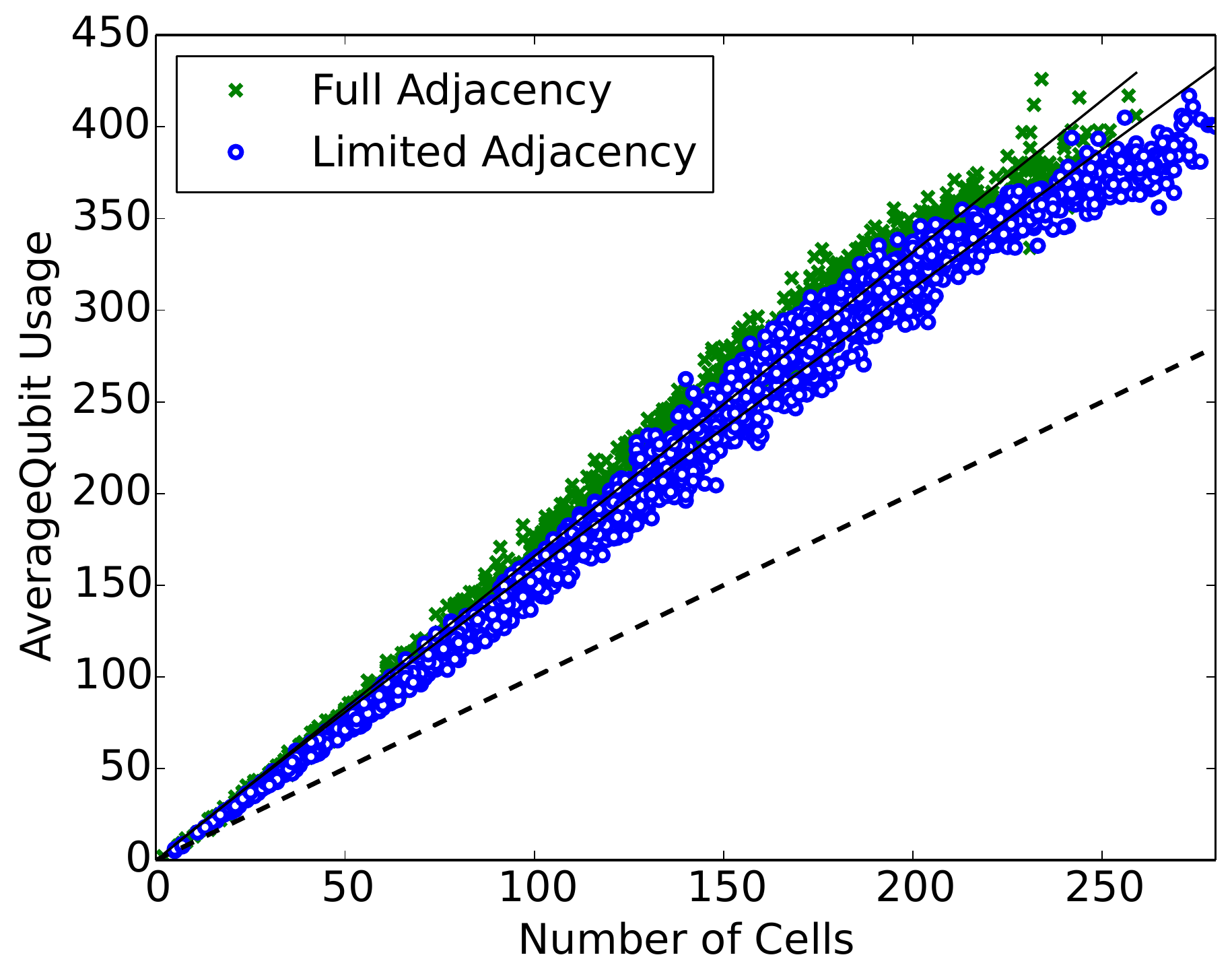}}
\caption{Average number of qubits needed to embed generated QCA circuits of different sizes. The solid black lines show power function fits.}
\label{fig: gen-qbits}
\end{figure}

\begin{figure}
\centering
\newcommand{\WW}{.81\columnwidth}
\subfloat[Dense Placement Algorithm]{\includegraphics[width=\WW]{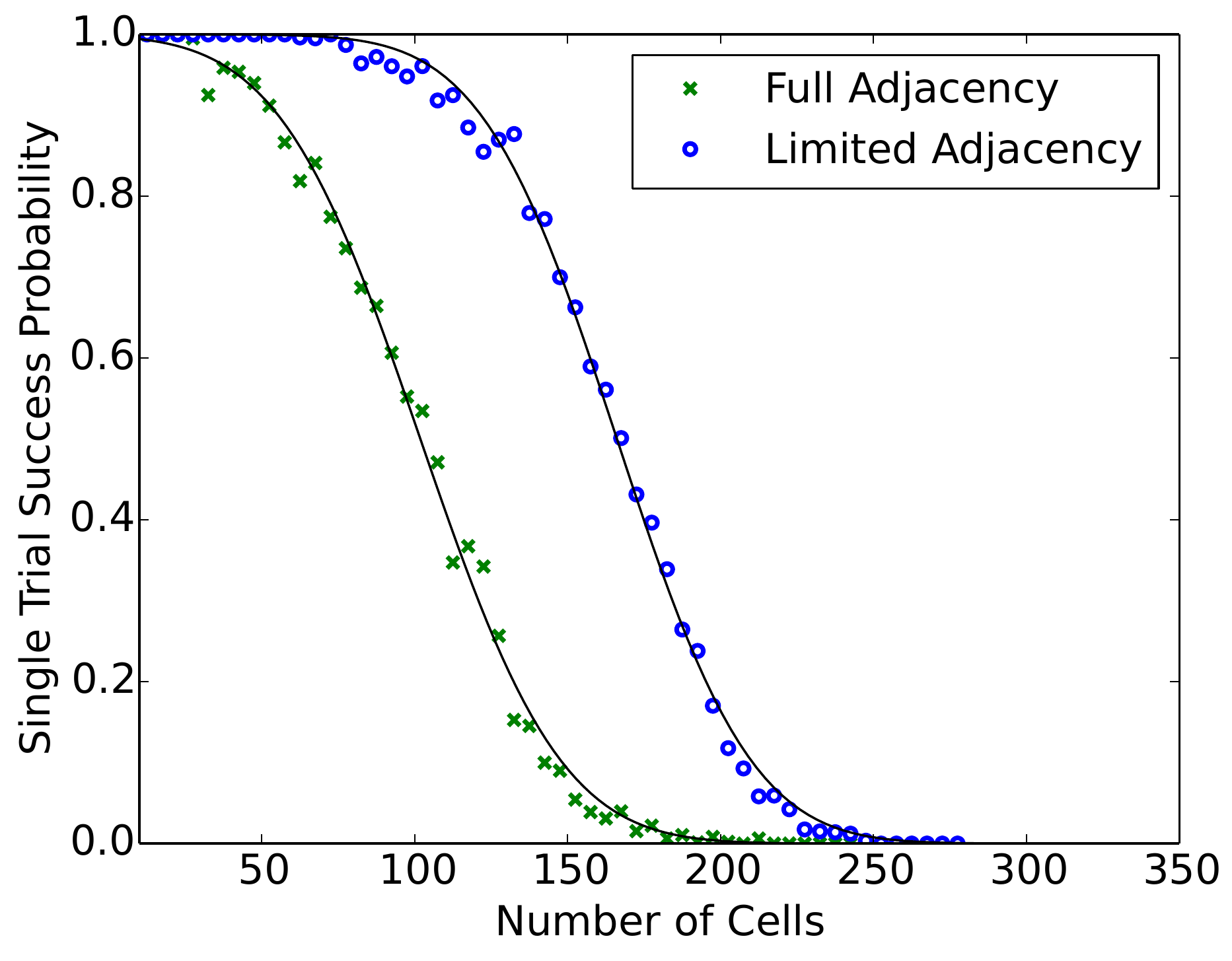}}\\
\subfloat[Heuristic Algorithm]{\includegraphics[width=\WW]{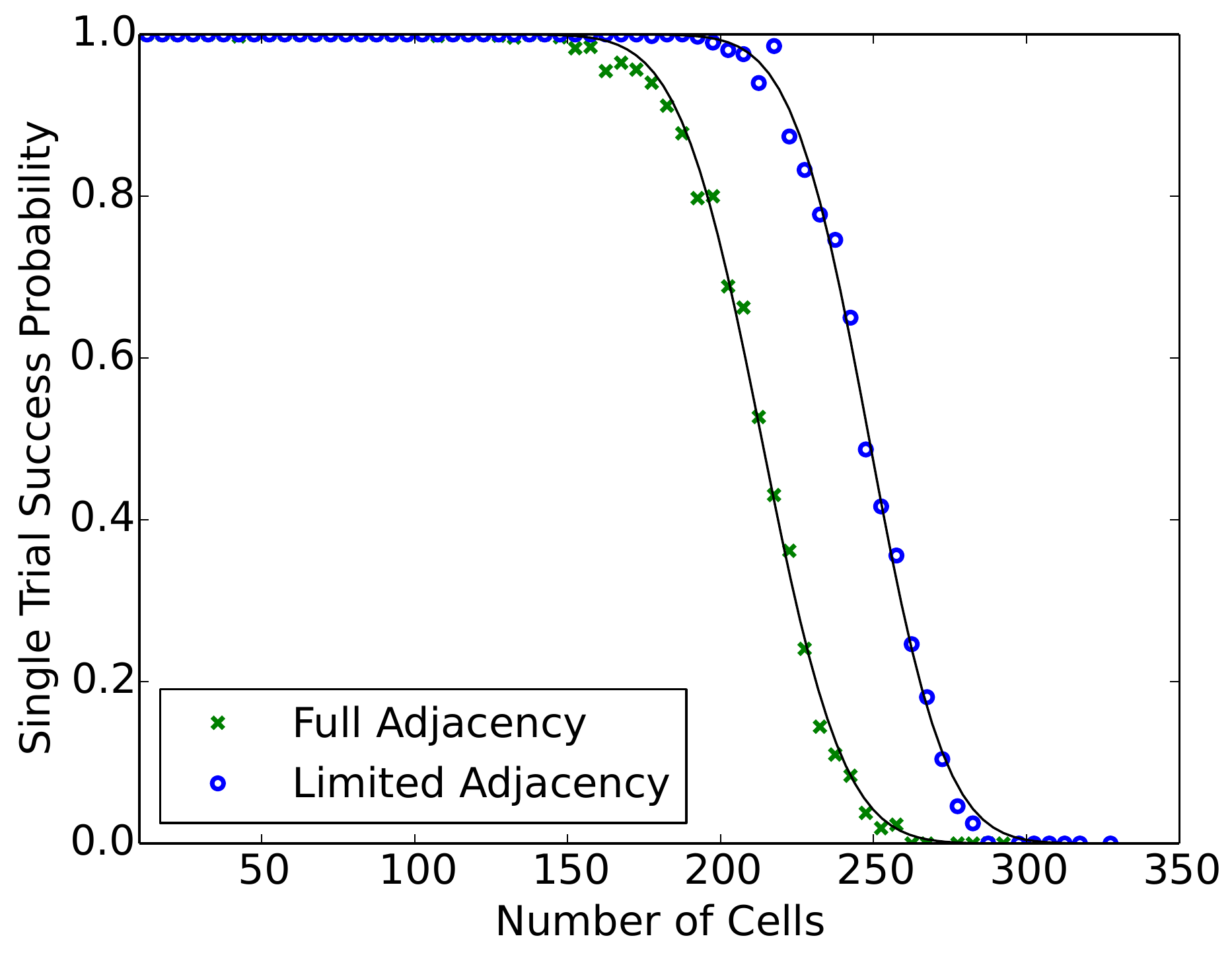}}
\caption{Average single trial embedding success probabilities using both algorithms. Points are obtained by binning the generated circuits by size. Solid lines show complementary error function fits.}
\label{fig: gen-probs}
\end{figure}

To understand what range of QCA circuits can be embedded within D-Wave's architecture, it was necessary to implement a stochastic circuit generator which constructs QCA circuits with random numbers of primitive circuit components and connecting wire lengths. The generated circuits were made exclusively of inverters, majority gates, driver cells, and wires. Connecting wires are generated in the connectivity graph representation as random trees with the root at the output of a primitive component and each leaf at an input of another primitive component.

We generated 2000 circuits for both limited and full adjacency and attempted to find embeddings within a 512 qubit processor using both algorithms. Ten trials were run for each circuit with a maximum allowable run-time per trial of 30 seconds. Fig. \ref{fig: gen-qbits} shows the average number of qubits used against the number of cells in the generated circuits for successful embeddings. Power function fits are included which indicate an approximately linear, or slightly super-linear in the case of the Dense Placement algorithm with limited adjacency, trend. As in the benchmark circuits, the Dense Placement algorithm performs better for limited adjacency than the Heuristic algorithm with the opposite true for full adjacency. Notably, the qubit usage in the benchmark circuit embeddings is consistent with the generated circuit data suggesting that the designed circuits are neither more nor less difficult to embed. The fall-off in qubit usage for the heuristic algorithm is likely a consequence of the circuit generator. Fig. \ref{fig: gen-probs} shows the average single trial embedding success probability as a function of the number of cells. It was observed that the success probability as a function of the number of cells, $N$, is approximately of the form 
\begin{equation}
p(N) = \frac{1}{2}\text{erfc}\left( \frac{N-\mu}{\Delta} \right)
\end{equation}
with $\text{erfc}$ the complementary error function, $p(\mu) = 0.5$, and $\Delta$ a measure of the width of the decline about $\mu$. Fitted parameters are included in Table \ref{table:512-parameters}. It is clear that even though the Dense Placement algorithm uses fewer qubits in limited adjacency, the Heuristic algorithm is more successful at embedding larger circuits.

\begin{table}
\caption{Fit parameters for the embedding success probability of generated circuits for a 512 qubit processor with $2\sigma$ errors for both Dense Placement (DP) and Heuristic (Heur) algorithms for limited (L) and full (F) adjacencies.}
\label{table:512-parameters}
\centering
\begin{tabular}{|c||c|c|}
\hline
\bfseries Method & $\mu$ (cells)	& $\Delta$ (cells) \\
\hline \hline
DP-L		&	$166.3 \pm 0.7$	&	$49.2 \pm 1.4$ \\
DP-F	&	$101.9 \pm 0.8$	&	$51.2 \pm 1.7$ \\
\hline
Heur	-L	&	$ 248.3\pm 0.4$	&	$29.8 \pm 0.8$ \\
Heur-F	&	$213.4 \pm 0.3$	&	$18.8 \pm 0.6$ \\
\hline
\end{tabular}
\end{table}

\begin{figure}
\centering
\newcommand{\WW}{.8\columnwidth}
\subfloat[Dense Placement Algorithm]{\includegraphics[width=\WW]{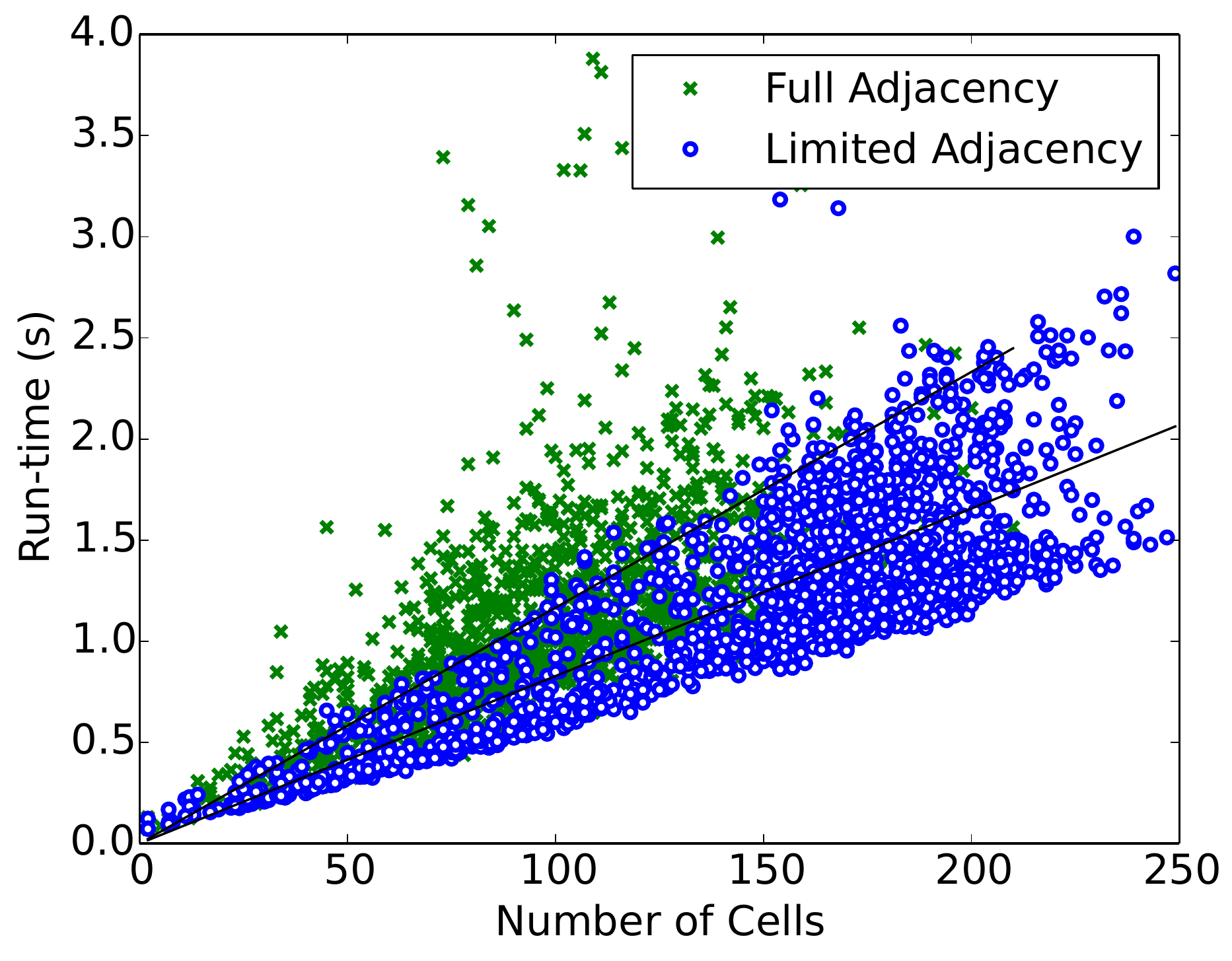}}\\
\subfloat[Heuristic Algorithm]{\includegraphics[width=\WW]{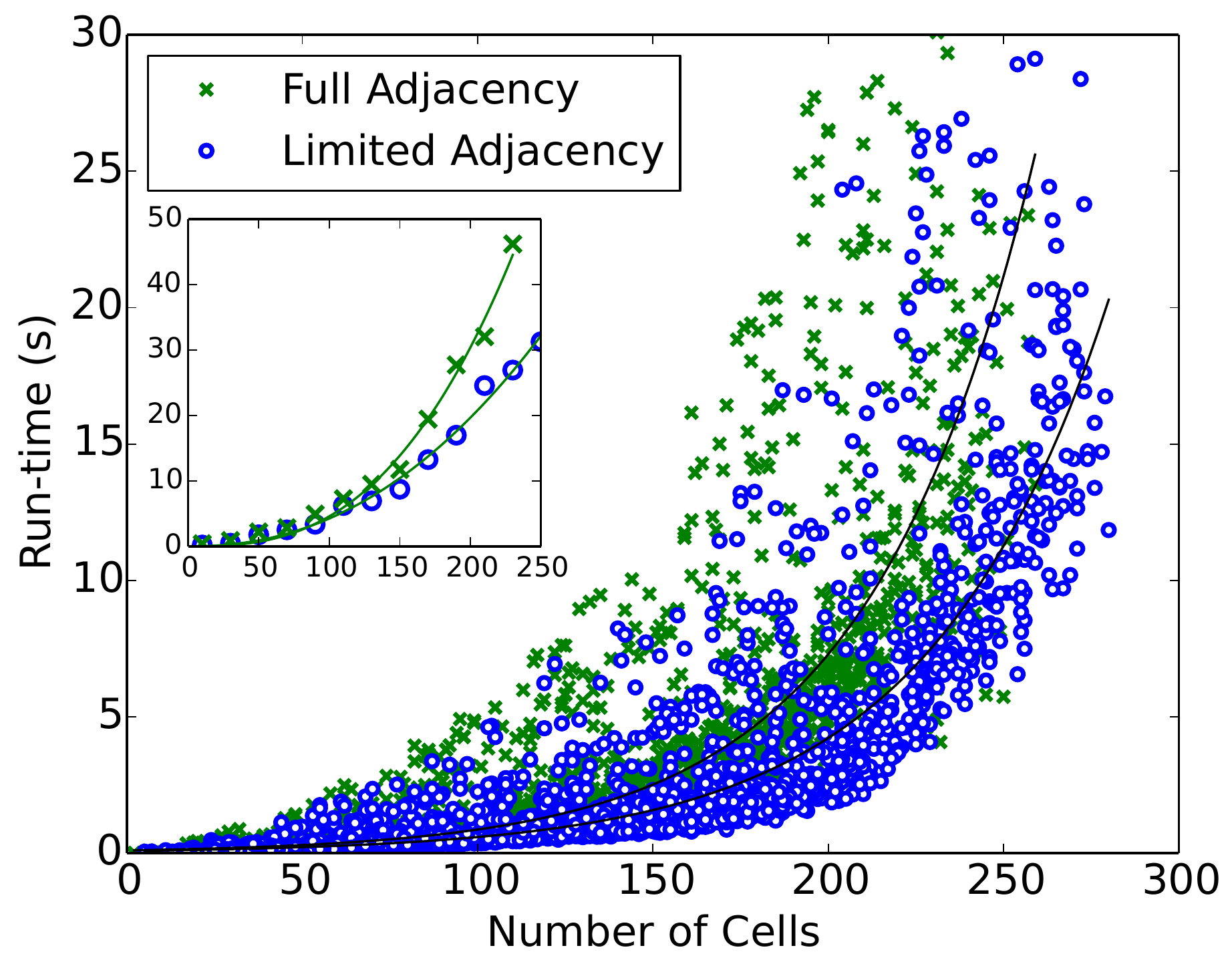}}
\caption{Average run-times for generated circuits on the 512 qubit processor for both algorithms with best fits. The inset of (b) shows the maximum run-times for bins of 20 cells with a power fit.}
\label{fig: 512-runtimes}
\end{figure}

We have not yet commented on embedding run-times. For such a specific class of embedding problems, it is not meaningful to discuss theoretical worst-cases for time complexity. Instead,  we consider only a best-fit to the average run-times for the generated circuits. More specifically, we quantize the algorithm run-times by the value of the fitted model at the 50\% embedding probability (50P) size $\mu$. Fig. \ref{fig: 512-runtimes} shows the average run-times for the generated circuits for both algorithms and both adjacency types. For the Dense Placement algorithm, the local nature of cell placement leads one to expect an approximately linear relationship between circuit size and run-time, as can be seen. For the Heuristic algorithm run-times, the best fit model was found to be exponential. Note the significant difference in time scale between the two algorithms, particularly for large circuits. The Heuristic algorithm is reported in \cite{cai2014} to have a worst case time complexity of $O(n_Hn_Ge_H(e_G + n_G\log(n_G))$ with $n$ and $e$ the number of nodes and edges in the source graph, $H$, and the target graph, $G$. In our case $H$ is the QCA circuit connectivity graph and $G$ the Chimera graph. Depending on the complexity of a circuit of $N_{cells}$ cells and the range of included interactions, the number of edges in $H$ is between $N_{cells}-1$ and $\frac{1}{2}N_{cells}(N_{cells}-1)$. Hence the worst case time complexity of the heuristic algorithm should be between $O(N_{cells}^2N_{qbits}^2\log(N_{qbits}))$ and $O(N_{cells}^3N_{qbits}^2\log(N_{qbits}))$ for a processor containing $N_{qbits}$ qubits. For a given processor, we should then expect the run-times to have an upper bound of the form $t_{max}(N) \propto N^\alpha$ with $2 \leq \alpha \leq 3$. The inset in Fig. \ref{fig: 512-runtimes}b shows the maximum run-times within bins of 20 cells and the associated power law fits. The trends for limited and full adjacency were found to have powers of $2.21 \pm .16$ and $2.73 \pm .21$ which lie within expected values. All embeddings were done using a single core of an Intel Core i7-2670QM processor.

\subsection{Scaling performance}
\label{subsec: scaling-performance}

\begin{figure}
\centering
\newcommand{\WW}{.8\columnwidth}
\subfloat[Embeddable circuit size, $\mu$]{\includegraphics[width=\WW]{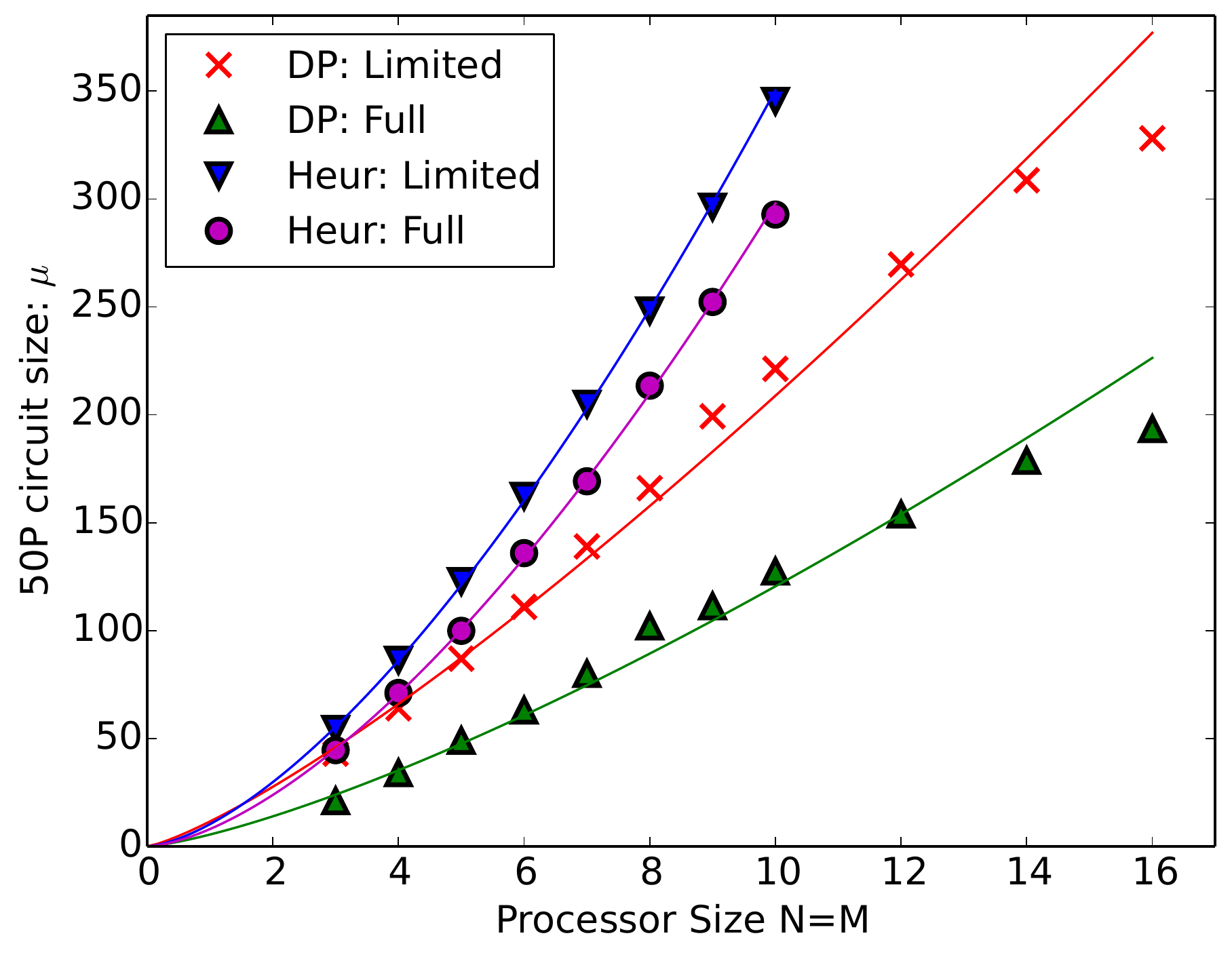}}\\
\subfloat[Run-time metric, $t(\mu)$. The inset has a reduce y-axis range.]{\includegraphics[width=\WW]{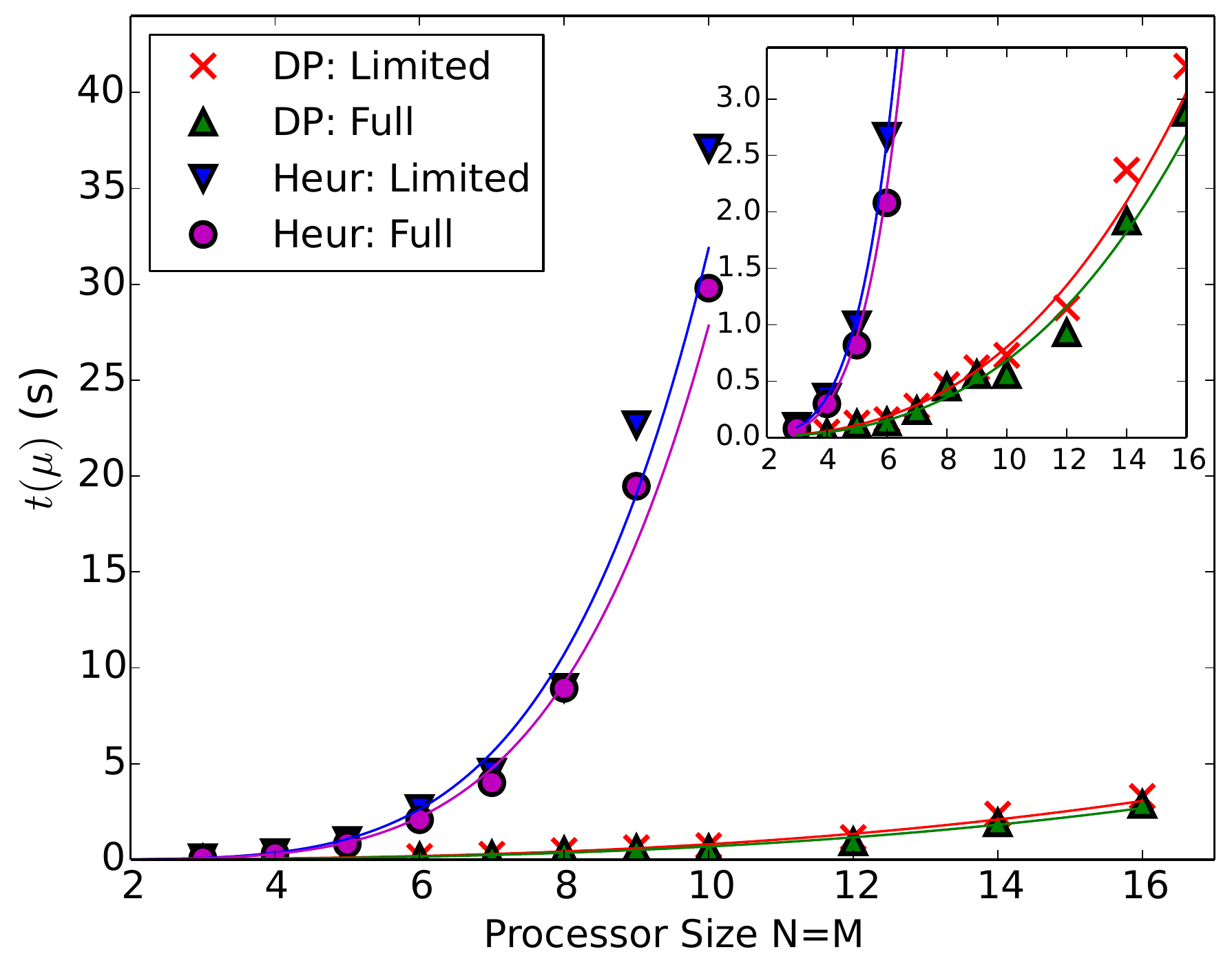}}
\caption{Metrics for the 50P circuit size and characteristic run-time for both algorithms and adjacencies as a function of the processor size.}
\label{fig:vs-N}
\end{figure}

\begin{table}
\centering
\caption{Power law fit parameters for the characteristic circuit size, $\mu$, and run-time $t(\mu)$.}
\begin{tabular}{|c||c|c||c|c|}
\hline
\bfseries Method		& $A_{\mu}$ (cells)	& $b_\mu$	&	$A_t$ (ms)	& $b_t$ \\
\hline \hline
DP-L	& $11.6 \pm 1.0$	&	$1.26 \pm 0.04$	&	$1.15 \pm 0.20$	&	$2.85 \pm 0.08$\\
DP-F	& $5.5 \pm 0.6$	&	$1.34 \pm 0.06$	&	$0.84 \pm 0.19$	&	$2.91 \pm 0.11$\\
\hline
Heur-L	& $10.3 \pm 0.2$	&	$1.53 \pm 0.01$	&	$0.41 \pm 0.11$	&	$4.89 \pm 0.14$\\
Heur-F	& $8.0 \pm 0.2$	&	$1.57 \pm 0.02$	&	$0.31 \pm 0.06$	&	$4.95 \pm 0.11$\\
\hline
\end{tabular}
\label{table:nm-data}
\end{table}

Here we investigate the trend of the 50P circuit size, $\mu(N)$, and the characteristic run-time, $t(\mu, N)$, of both algorithms over a range of processor sizes. We consider only square processors. That is, we assume there are $N$ rows and $M = N$ columns of $8$ qubit tiles. Due to its exponential run-time performance, we only tested the Heuristic algorithm up to processors of size $N=10$. The Dense Placement algorithm was tested on processors up to size $N=16$. The estimated values of $\mu(N)$ and associated run-times $t(\mu, N)$ are shown in Fig. \ref{fig:vs-N} with power law fits. The fit parameters are included in Table \ref{table:nm-data} where $\mu(N) = A_\mu N^{b_\mu}$ and $t(\mu, N) = A_tN^{b_t}$. The inset in Fig. \ref{fig:vs-N}b has the y-range expanded to better see the Dense Placement algorithm performance. While the power law fits for $\mu(N)$ appear accurate up to $N=10$, there is a notable fall-off for the  Dense Placement algorithm for larger processor sizes, likely due to the need to coordinate the discussed long chains for the final cell placements.

\subsection{Resilience to Processor Yield}
\label{subsec: proc-yield}

\begin{figure}
\centering
\includegraphics[width=.8\columnwidth]{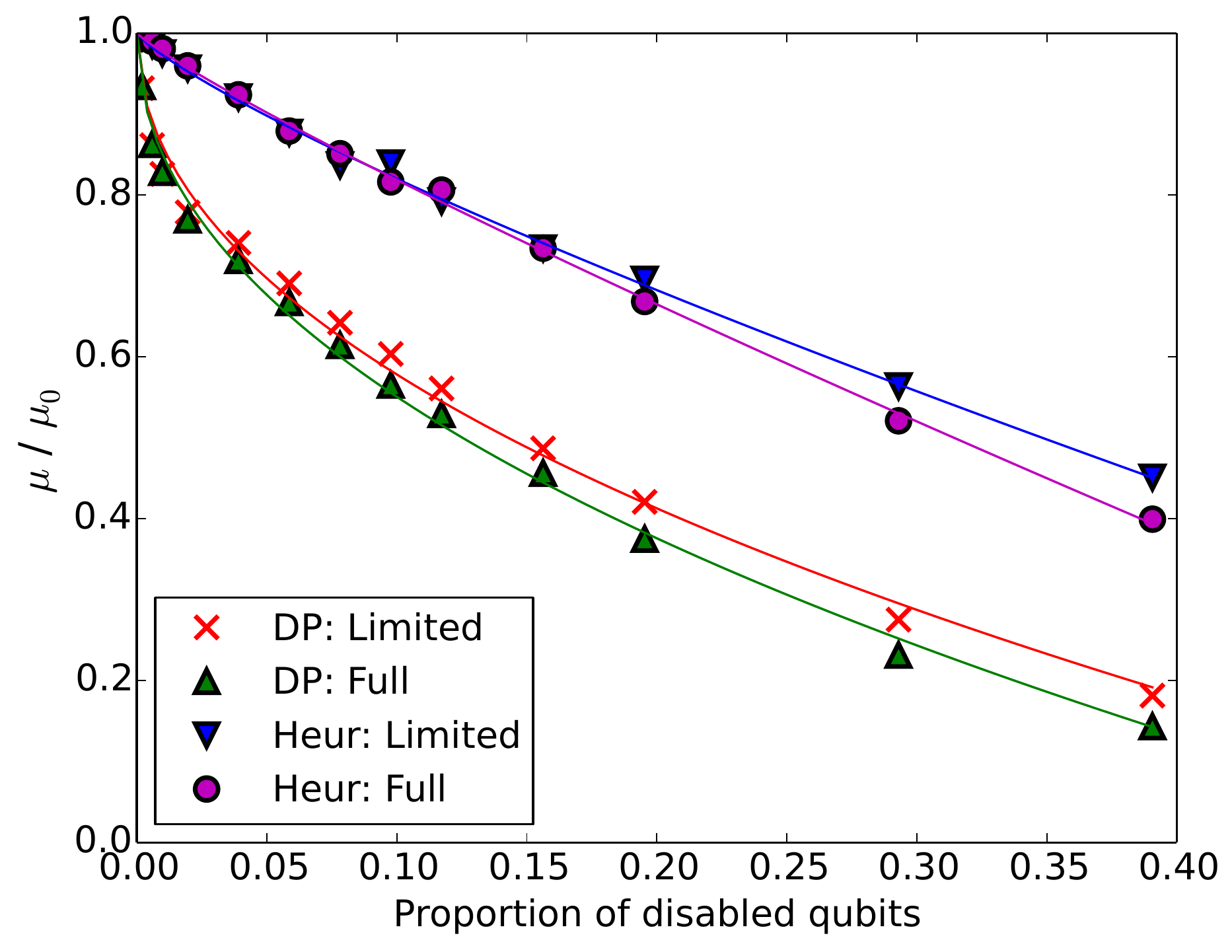}
\caption{Relative decrease in the embeddable circuit size as a function of the proportion of qubits disabled. Fits are shown.}
\label{fig:vs-ndis}
\end{figure}

So far, we have assumed all of the qubits and couplers in the processor to be available/active. Here we investigate the performance of both algorithms for generated circuits with different percentages of the available qubits disabled. A new set of disabled qubits is uniformly generated for each embedding trial for each circuit. Fig. \ref{fig:vs-ndis} shows the relative $\mu$ values for different percentages of disable qubits on a 512 qubit processor. Here $\mu_0$ indicates the value of $\mu$ when no qubits are disabled. We fit a model of the form
\begin{equation}
\mu(n_{dis}) = \mu_0 \left[1-\alpha n_{dis}^\beta \right]
\end{equation}
where $0 \leq n_{dis} \leq 1$ is the proportion of qubits disabled. As typical values of $n_{dis}$ are small ($\sim 0.05$), of importance is the observation that $\mu(n_{dis})$ for the Dense Placement algorithm falls off quickly with $n_{dis}$. In contrast, the Heuristic algorithm seems to be much more resilient to yield.

\section{Conclusion}
\label{sec:conclusion}

In this paper, we have discussed the potential for applying D-Wave Systems Inc.'s current quantum annealing architecture to investigate characteristics of large QCA circuits. We have introduced two approaches to the problem of embedding QCA circuits onto D-Wave's quantum annealing processor and characterised these approaches using benchmark and generated circuits. Estimates were made of the embeddable circuit sizes and characteristic run-times for different processor sizes, with characteristic circuit sizes defined as the size at which a single embedding attempt was 50\% successful. The Dense Placement algorithm, efficient at tightly packing low connectivity nodes, required fewer qubits to embed QCA circuits when a limited set of interactions was considered. For a higher connectivity representation of QCA circuits, D-Wave's Heuristic algorithm required fewer qubits. In all cases, the Heuristic algorithm was able to embed larger circuits than the Dense Placement algorithm. However, the local nature of the Dense Placement algorithm allowed for significantly faster run-times. It was observed that the embeddable circuit size for the Heuristic algorithm was more resilient to disabled qubits than the Dense Placement algorithm. For the the 512 qubit Vesuvius processor with no disabled qubits, we estimate characteristic circuit sizes of approximately 166 and 102 cells for limited and full adjacency for the Dense Placement algorithm and 248 and 213 cells respectively for the Heuristic algorithm with characteristic run-times for the two algorithms of approximately 0.5 and 9 seconds. For the 1152 qubit ``Washington'' processor, we estimate characteristic circuit sizes of 270 and 154 cells for limited and full adjacency with the Dense Placement algorithm and, extending the observed trend for the Heuristic algorithm, approximately 460 and 400 cells respectively for the Heuristic algorithm with characteristic run-times of approximately 1 and 80 seconds.


\bibliographystyle{IEEEtran}
\bibliography{\BIBFILE}

\end{document}